\documentclass[%
 reprint,
superscriptaddress,     
 amsmath,amssymb,
 aps,
]{revtex4-2}

\usepackage{graphicx}
\usepackage{dcolumn}
\usepackage{bm}

\mathchardef\mhyphen="2D
\def\mm{\scalebox{0.9}{-}}
\def\pp{\scalebox{0.6}{+}}

\begin{document}

\preprint{APS/123-QED}

\title{Time-Varying Materials in Presence of Dispersion: Plane-Wave Propagation in a Lorentzian Medium with Temporal Discontinuity}

\author{Diego M. Sol\'{i}s}
\affiliation{Department of Electrical and Systems Engineering, University of Pennsylvania, Philadelphia, Pennsylvania 19104, USA}
\author{Raphael Kastner}
\affiliation{Department of Electrical and Systems Engineering, University of Pennsylvania, Philadelphia, Pennsylvania 19104, USA}
\affiliation{Tel Aviv University, Tel Aviv, 69978 Israel}
\author{Nader Engheta}
\email[Corresponding author: ]{engheta@seas.upenn.edu}
\affiliation{Department of Electrical and Systems Engineering, University of Pennsylvania, Philadelphia, Pennsylvania 19104, USA}


\date{\today}

\begin{abstract}
We study the problem of a temporal discontinuity in the permittivity of an unbounded medium with Lorentzian dispersion. More specifically, we tackle the situation in which a monochromatic plane wave forward-travelling in a (generally lossy) Lorentzian-like medium scatters from the temporal ``half-space interface'' that results from an abrupt temporal change in its plasma frequency (while keeping its resonance frequency constant). In order to achieve momentum preservation across the temporal discontinuity, we show how, unlike in the well-known problem of a nondispersive discontinuity, the second-order nature of the dielectric function now gives rise to two shifted frequencies. As a consequence, whereas in the nondispersive scenario the continuity of the electric displacement $\mathbf{D}$ and the magnetic induction $\mathbf{B}$ suffice to find the amplitude of the new forward and backward wave, we now need two extra temporal boundary conditions. That is, two forward and two backward plane waves are now instantaneously generated in response to a forward-only plane wave. We also include a transmission-line equivalent with lumped circuit elements that describes the dispersive time-discontinuous scenario under consideration.
\end{abstract}

\maketitle


\section{\label{sec:level1}Introduction}
In the past few years, time-variant metamaterials/metasurfaces have become a hot research topic within the photonics community, given their potential to boost the degree of manipulation of light-matter interactions achieved by their time-invariant predecessors. The latter, through the subwavelength space-modulation of the electric and/or magnetic response \cite{engheta2006metamaterials}, allow for alluring possibilities in the way light is controlled, thus enabling a vast range of interesting phenomena and promising applications, from strengthened nonlinearities \cite{PhysRevLett.102.043907} and $\epsilon$-near-zero (ENZ) propagation \cite{PhysRevLett.97.157403,PhysRevLett.100.033903} to artificial Faraday rotation \cite{doi:10.1063/1.3615688} and optically-driven topological states \cite{Gorlach2018}. On the other hand, an externally-induced time-modulation in some of the properties of these engineered structures largely broadens the degree of harnessing of light manipulation, in which case we have a time-varying metamaterial. This spatio-temporal modulation is the supporting platform of such fascinating effects as magnetless nonreciprocity \cite{Yu2009} or time reversal \cite{Bacot_2016}, just to name a few. In this regard, the research on active metasurfaces has gained a lot of momentum in the past few years \cite{Shaltouteaat3100,https://doi.org/10.1002/adma.201606422,Lee2018,https://doi.org/10.1002/adma.201904069}.

One avenue to induce this temporal variation is the time-modulation of a medium's dielectric function, e.g. electro-optically. In \cite{PhysRevLett.32.1101}, a nonstationary interface was reported from plasma ionization by a high-power electromagnetic pulse. The problem of wave propagation in an unbounded medium with a rapid change---and, to a lesser extent, a slab with sinusoidal time variation---in its constitutive parameters was first theoretically studied in \cite{1124533} for the case of nondispersive permittivity and/or permeability. These nondispersive step transients, further explored in \cite{Xiao:14,8858032}, effectively produce a ``time interface'': based on the continuity of $\mathbf{D}$ and $\mathbf{B}$, an instantaneous frequency shift occurs to accommodate the new permittivity while preserving the wave momentum, and a forward and a backward wave arise whose amplitudes are quantified by what can be seen as the temporal dual of the Fresnel coefficients. These step-like discontinuities were later analyzed e.g. in a half-space \cite{1139931} and a dielectric layer \cite{Fante1973}. Moreover, \cite{Xiao:11,Xiao:11b} addressed the adiabatic frequency conversion of optical pulses going through slabs with arbitrarily time-varying refractive index, while \cite{HAYRAPETYAN2016158} and \cite{chegnizadeh2018general} reported wave solutions for a smooth or arbitrary transition of the refractive index, respectively. Wave propagation undergoing periodic temporal inhomogeneities of the permittivity has also been investigated in a half-space \cite{86907}, a slab \cite{1138637,7766053,PhysRevA.79.053821,Zurita-Sanchez:12,PhysRevA.96.063831,PhysRevA.98.053852}, or a spacetime-periodic (traveling-wave modulation) medium \cite{1124657,1125340,5252018,Fante:72}: time-periodic variations exhibit frequency-periodic band-structured dispersion relations that include wavevector gaps \cite{PhysRevA.79.053821}, dual of the bandgaps of space-periodic media. As shown in \cite{1138637,7766053,PhysRevA.98.053852,8434236,PhysRevA.97.013839,PhysRevLett.120.087401,9103963}, this time-Floquet modulation can be harnessed to achieve parametric amplifiers.

Nonetheless, most of the aforementioned works consider nondispersive susceptibilities only (excepting \cite{1139931,Fante1973}, where a plasma is parameterized with a nonstationary electron density, and \cite{86907}, where the time-varying parameter is conductivity). In \cite{1139657}, on the contrary, closed-form Green's functions are obtained for pulsed excitations within spatially homogeneous media with abrupt or gradual temporal changes, either without dispersion or considering a cold ionized lossless plasma described with Drude dispersion. Very recently, the question of time-varying dispersion has been studied from different angles; namely, a transmission line \cite{PhysRevApplied.11.014024} and a meta-atom \cite{PhysRevResearch.1.023014} with time-modulated reactive loads, and the analysis of the instantaneous radiation of nonharmonic dipole moments \cite{PhysRevA.102.013503} and nonstationary Drude-Lorentz polarizabilities \cite{mirmoosa2020dipole}.

In the present work, we assume an initial plane wave at $t\!<\!0$ with frequency $\omega_{\mm}$ and bring in the effects of Lorentzian dispersion when considering a step-like change in the plasma frequency $\omega_p$ with otherwise constant resonance frequency $\omega_0$. Unlike in \cite{1124533,Xiao:14,8858032}, this abrupt change gives rise to two shifted frequencies (in the simplified lossless case, a lower frequency $\omega_1\!<\!\omega_0$ and an upper frequency $\omega_2\!>\!\omega_0$) that bear the following interpretation when $\omega_0$ is considerably larger than $\omega_{\mm}$: while $\omega_1$ reflects in essence the change in permittivity similarly to the nondispersive case, $\omega_2$ characterizes a wave of a different nature, viz. one that has negligible magnetic component; the medium at $\omega_2$ thus possesses ENZ characteristics.

We begin by defining in Sec.~\ref{sec:Lorentzian} the differential equation describing the Lorentzian-like dielectric response characterizing our time-varying medium to further derive the initial conditions across the temporal change in $\omega_p$ at $t\!=\!0$. This transition is perceived as abruptly varying the volumetric density of $\omega_0$-resonating dipoles $N\!=\!N(t)$, our control parameter. As a starting point, we mainly look into the case where this number changes from zero to a specified value $N_{\pp}$. From the differential equation relating the polarization vector $\mathbf{P}$ to the electric field $\mathbf{E}$, we show that $\mathbf{E}$, $\mathbf{P}$ and $\frac{d\mathbf{P}}{dt}$ are all continuous across the temporal discontinuity at $t\!=\!0$. In Sec.~\ref{sec:momentum} we use preservation of momentum to analytically find $\omega_1$ and $\omega_2$ and also give a detailed numerical account for the evolution of the frequency split over time when the transition is gradual rather than abrupt. A dynamic analysis towards a full wave solution is developed in Sec.~\ref{sec:dynamics} for a lossless scenario. The approach is firstly based, in Sec.~\ref{subsec:dynamicsA}, on the scattering-parameter model from \cite{8858032}. It is further substantiated---and confirmed---in Sec.~\ref{subsec:dynamicsB} by a Laplace-transform-based first-principles solution to the amplitudes for the forward and backward propagating constituents at $\omega_1$ and $\omega_2$; this comprehensive development also recovers $\omega_1$ and $\omega_2$. Furthermore, we developed a finite-difference time-domain (FDTD) \cite{taflove2005computational} solver whose simulation results perfectly agree with our analytical predictions. In Sec.~\ref{sec:transmission_line}, we show how this one dimensional spatial problem may be likened to a transmission line equivalent that is relatively simple to use. Further phenomena related to losses are described In Sec.~\ref{sec:lossy}. Finally, conclusions are drawn in Sec.~\ref{sec:conclusions}.

\section{Time-Varying Lorentzian Dispersion: Initial Conditions}\label{sec:Lorentzian}
\newcommand{\comment}[1]{}
Let us consider, for $t\!<\!0$, an $\widehat{\mathbf{x}}$-polarized electric field plane wave traveling in the $+z$ direction and oscillating at a purely real frequency $\omega_{\mm}$ in an unbounded dispersive medium (for simplicity, we will assume it lossless for now) whose electric polarization charge $P$ responds to the electric field $E$ following a susceptibility $\chi$ that can be described in the frequency domain by a Lorentzian resonance centered at $\omega_0$, such that
\begin{equation}
P(\omega)=\epsilon_0\chi(\omega)E(\omega)=\epsilon_0\frac{\omega_p^2}{\omega_0^2-\omega^2}E(\omega),\label{eq:200}
\end{equation}
where $\omega_p$=$q_e\sqrt{\frac{N}{\epsilon_0 m_e}}$ is the plasma frequency, $N$ being the volumetric density of polarizable atoms, with $m_e$ and $q_e$ the electron's mass and charge, respectively. In the time domain, this relation adopts the form of the following second-order differential equation
\begin{equation}
\frac{d^2P(t)}{dt^2}+\omega_0^2P(t)=\epsilon_0\omega_p^2E(t),\label{eq:201}
\end{equation}
which can be also written as the convolution $P(t)$=$\epsilon_0\chi(t)\underset{t}{*}E(t)$, where $\chi(t)=\frac{\omega_p^2}{\omega_0}\text{sin}\left(\omega_0t\right)U(t)$ is the system's impulse response, $U(t)$ being the step function, and $\underset{t}{*}$ denoting the linear time-invariant (LTI) convolution operation with respect to $t$. 

Now, let us allow for $N$---and thus $\omega_p$---to be time-dependent, and consider a scenario where it abruptly changes at $t\!=\!0$ as $N(t)\!=\!N_{\mm}\!+\!(N_{\pp}\!-\!N_{\mm})U(t)$, with $N_{\mm}$ and $N_{\pp}$ some arbitrary positive constants. After defining $A(t)\!=\!\omega_p^2(t)$, Eq.~\eqref{eq:201} becomes
\begin{equation}
\frac{d^2P(t)}{dt^2}+\omega_0^2P(t)=\epsilon_0A(t)E(t),\label{eq:202}
\end{equation}
where we have indicated that the resonance frequency $\omega_0$ is time-invariant. Moreover, despite our linear system now being time-variant (LTV), one can still use the convolution operator and write $P(t)$=$\epsilon_0\chi_n(t)\underset{t}{*}\big(A(t)E(t)\big)$ \cite{solis2020generalization}, where we have defined the normalized impulse response $\chi_n(t)\!=\!\frac{1}{\omega_0}\text{sin}\left(\omega_0t\right)U(t)$, or, in the frequency domain,
\begin{equation}
P(\omega)=\epsilon_0\frac{\frac{1}{2\pi}A(\omega)\underset{\omega}{*}E(\omega)}{\omega_0^2-\omega^2},\label{eq:203}
\end{equation}
with $\chi_n(\omega)\!=\!\frac{1}{\omega_0^2-\omega^2}$. In short, the dielectric response to an impulse applied at time $\tau$ is only a function of $N(t\!=\!\tau)$ and not of $N(t\!>\!\tau)$: intuitively, the new dipoles brought into the medium after the electric-field impulse at $t\!=\!\tau$ simply have no excitation to respond to; mathematically, this can be traced back to the invariance of the coefficients in the left-hand side of Eq.~\eqref{eq:202}, and gives us one key piece of information: regardless of the step-function discontinuity in $N(t)$, $P(t)$ is continuous (note that $\chi_n(\!t=\!0)\!=\!0$), and so is $\frac{dP(t)}{dt}$ (only a spike in $E$ would determine otherwise). In the more general framework of LTV systems, the response observed at time $t$ due to an impulse at time $\tau$ can in this case be recast as $h(t,\tau)\!=\!\epsilon_0\!A(\tau)\chi_n(t-\tau)$, which allows us to write
\begin{equation}
P(t)=\int_{-\infty}^{t}h(t,\tau)E(\tau)d\tau.\label{eq:204}
\end{equation}
Importantly, the depicted situation differs from the model assumed in \cite{mirmoosa2020dipole}, where $h(t,\tau)\!=\!\epsilon_0\!A(t)\chi_n(t-\tau)$. Formally, our continuity of both $P(t)$ and $\frac{dP}{dt}$ across $t\!=\!0$ can be substantiated as follows. Applying the one-sided Laplace transform $\mathcal{L} \left\lbrace f(t)\right\rbrace\!=\!\tilde{f}(s)$, defined over the temporal interval $t\!:\![0^{\mm},\infty)$, to Eq.~\eqref{eq:202}, and solving for $\tilde{P}(s)$, we have
\begin{equation}\label{eq:motiontime_31}
\tilde{P}(s)=\frac{\epsilon_0\mathcal{L}\left\lbrace A(t)E(t) \right\rbrace +sP(0^{\mm})+	\frac{dP}{dt}(0^{\mm})}{s^2+\omega_0^2},
\end{equation}
where, e.g., $P(0^{\mm})$ stands for $P(t\!=\!0^{\mm})$. A direct application of the initial value theorem (IVT) \cite{oppenheim1996s}
\begin{equation}\label{eq:infty}
P(0^{\pp})=\lim\limits_{s\to\infty}s\tilde{P}(s)
\end{equation}
provides the continuity condition for $P$:
\begin{equation}\label{eq:inital_P}
P(0^{\pp})=P(0^{\mm}).
\end{equation}
Similarly, for  $\frac{dP}{dt}$,
\begin{equation}\label{eq:infty_der}
\frac{dP}{dt}(0^{\pp})=\lim\limits_{s\to\infty}\left[ s^2\tilde{P}(s)-sP(0^{\pp})\right] .
\end{equation}
However, by virtue of Eq.~\eqref{eq:inital_P} and substituting Eq.~\eqref{eq:motiontime_31} with the understanding that $\lim\limits_{s\to\infty}\mathcal{L}\left\{A(t)E(t) \right\}\!=\!0$, we find that $\frac{dP}{dt}$ is continuous as well:
\begin{equation}\label{eq:dpdt}
\frac{dP}{dt}(0^{\pp})=\lim\limits_{s\to\infty}\left[ s^2\tilde{P}(s)\!-\!sP(0^{\mm})\right]
=\frac{dP}{dt}(0^{\mm}).
\end{equation}
Finally, the Laplace-domain polarization emerges when $N_{\mm}\!=\!0$ as
\begin{equation}\label{eq:motiontime_4}
\tilde{P}(s)=\frac{\epsilon_0\mathcal{L}\left\lbrace A(t)E(t) \right\rbrace}{s^2+\omega_0^2},
\end{equation}
which immediately connects with Eq.~\eqref{eq:203}.

\section{Kinematics: Preservation of Momentum}\label{sec:momentum} 
The existence of dispersion does not change the fact that, as dictated by electromagnetic momentum conservation, the new waves arising after the temporal boundary must be shifted in frequency with respect to $\omega_{\mm}$, as shown in \cite{1124533,Xiao:14,8858032} for a nondispersive scenario. Our initial wave oscillating at $\omega_{\mm}$ has a wavenumber $k_{\mm}(\omega_{\mm})$ so, after the temporal jump, the supported new frequencies $\omega_{\pp}$ will be those that satisfy the equality $k_{\mm}(\omega_{\mm})$=$k_{\pp}(\omega_{\pp})$. This leads, when there is no magnetic response, to the transcendental equation $\frac{\omega_{\mm}}{c}\sqrt{\epsilon_{\mm}(\omega_{\mm})}$=$\frac{\omega_{\pp}}{c}\sqrt{\epsilon_{\pp}(\omega_{\pp})}$, which in our lossless case can be written, when $\epsilon_{\infty}$=1 and thus the relative dielectric permittivity $\epsilon(\omega)\!=\!1\!+\!\chi(\omega)$, as:
\begin{equation}
\omega_{\mm}\sqrt{1+\frac{\omega_{p\mm}^2}{\omega_0^2-\omega_{\mm}^2}}=\omega_{\pp}\sqrt{1+\frac{\omega_{p\pp}^2}{\omega_0^2-\omega_{\pp}^2}}.\label{eq:2A1}
\end{equation}
Squaring both sides of Eq.~\eqref{eq:2A1} leads to a quartic polynomial equation in $\omega_{\pp}$ whose 4 roots determine the new frequencies for $t\!>\!0$:
\begin{equation}
\omega_{\pp}=\pm\sqrt{\frac{K\pm\sqrt{K^2-4\omega_0^2\omega_{\mm}^2(\omega_{\mm}^2-\omega_0^2)(\omega_{\mm}^2-\omega_0^2-\omega_{p\mm}^2)}}{2(\omega_{\mm}^2-\omega_0^2)}},\label{eq:2A2}
\end{equation}
which we will denote $\pm\omega_{1}$ and $\pm\omega_{2}$, with 
\begin{equation}
K=\omega_{\mm}^2\left(\omega_{\mm}^2+\omega_{p\pp}^2-\omega_{p\mm}^2\right)-\omega_{0}^2(\omega_{0}^2+\omega_{p\pp}^2).\label{eq:2A3}
\end{equation}
For definiteness, we will choose $K\!+\!\sqrt{}$ for $\pm\omega_{1}$ and $K\!-\!\sqrt{}$ for $\pm\omega_{2}$ such that $\omega_{1}\!<\!\omega_{2}$ ($\omega_{2}\!<\!\omega_{1}$) when $\vert\omega_{\mm}\vert\!<\!\omega_0$ ($\vert\omega_{\mm}\vert\!<\!\sqrt{\omega_0^2+\omega_{p\mm}^2}$): note that only in the interval $\omega_0\!<\!\vert\omega_{\mm}\vert\!<\!\sqrt{\omega_0^2+\omega_{p\mm}^2}$ of anomalous dispersion, which we will not address and for which $\epsilon_{\mm}\!<\!0$, do we get complex solutions---more precisely, purely real (imaginary) $\omega_{1}$ ($\omega_{2}$)---. Specializing Eq.~\eqref{eq:2A1} to the case $\omega_{p\mm}\!=\!0$ (the medium is vacuum for $t\!<\!0$), we have the following characteristic equation for $\omega_{\pp}$:
\begin{equation}\label{eq:Lorentz-7-6}
\omega_{\pp}^4-( \omega_{\mm}^2+ \omega_0^2+ \omega_{p\pp}^2)\omega_{\pp}^2+ \omega_0^2 \omega_{\mm}^2=0,
\end{equation}
and Eq.~\eqref{eq:2A2} reduces to
\begin{equation}
\omega_{\pp}=\pm\sqrt{\frac{\omega_{\mm}^2+\omega_{0}^2+\omega_{p{\pp}}^2\pm\sqrt{(\omega_{\mm}^2+\omega_{0}^2+\omega_{p{\pp}}^2)^2-4\omega_0^2\omega_{\mm}^2}}{2}},\label{eq:2A2B}
\end{equation}

\begin{figure}[h]
\includegraphics[width=3.4in]{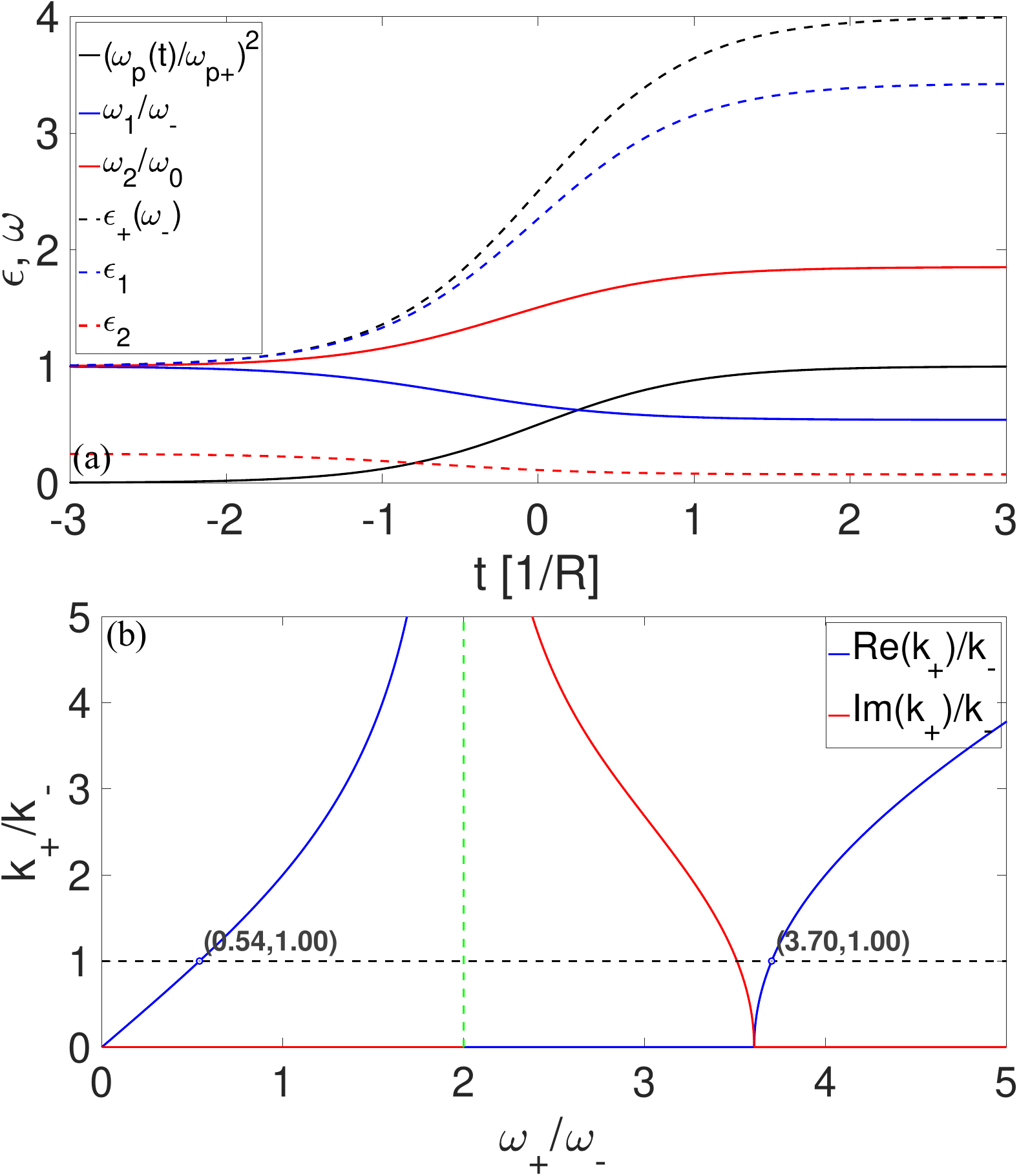}
\caption{(a) Temporal evolution of $\omega_l$ and $\epsilon_l$ ($l\!=\!1,2$) as $\omega_p(t)$ transitions from $\omega_{p\protect\mm}\!=\!0$ (vacuum) to $\omega_{p\protect\pp}$ such that $\epsilon_{\protect\pp}(\omega_{\protect\mm})\!=\!4$ with $\omega_0\!=\!2\omega_{\protect\mm}$, resulting in $\omega_{p\protect\pp}\!=\!3\omega_{\protect\mm}$. (b) Dispersion diagram $k_{\protect\pp}$ vs. $\omega_{\protect\pp}$, showing the two solutions for $\omega_{\protect\pp}$ that achieve momentum conservation. The dashed green line represents $\omega_0$.}
\label{fig:1}
\end{figure}

In order to illustrate how the frequencies evolve from $\omega_{\mm}$ to $(\omega_1,\omega_2)$, let us for a moment assume that $N(t)\!=\!N_{\mm}\!+\!(N_{\pp}\!-\!N_{\mm})\big(1\!+\!\text{tanh}(Rt)\big)/2$, with $R$ some constant describing the transition rate (in the limit $R\!\to\!\infty$, we have $\big(1\!+\!\text{tanh}(Rt)\big)/2\!\to\!U(t)$). In Fig.~\ref{fig:1}a, we consider a transition from vacuum ($N_{\mm}\!=\!0$) to a Lorentzian-like medium with $N_{\pp}$ chosen such that $\epsilon_{\pp}(\omega_{\mm})\!=\!4$, and with $\omega_0\!=\!2\omega_{\mm}$. As soon as $N_{\pp}\!>\!0$, $\omega_{\mm}$ splits into the pair $(\omega_1,\omega_2)\!=\!(\omega_{\mm},\omega_0)$. This is understood once we make $\omega_{p\mm}\!=\!0$ in Eq.~\eqref{eq:2A1} and take the limit $\omega_{p\pp}\!\to\!0$: in addition to the trivial solution $\epsilon_1(\omega_1\!=\!\omega_{\mm})\!=\!1$, we also have $\epsilon_2(\omega_2\!=\!\omega_0)\!=\!(\frac{\omega_{\mm}}{\omega_0})^2$, which in this case is equal to 0.25. Physically, this is just the manifestation of the natural frequency $\omega_0$ of the newly-added oscillators, which may surface depending on the boundary conditions. 

Now, we could think of ``quantizing'' $\text{tanh}(Rt)$ and consider the entire continuous transition $N_{\mm}\!\to\!N_{\pp}$ as a succession of infinitesimal step-function-like temporal-discontinuities. By doing this, we next go on by applying Eq.~\eqref{eq:2A1} twice in our second temporal jump, with both $\omega_{\mm}$ itself and $\omega_0$ as the input frequencies: it turns out that $\big(\omega_1(t),\omega_2(t)\big)$ are interrelated such that they both give rise to the same pair of output frequencies, so, notably, there is  a pair $\big(\omega_1(t),\omega_2(t)\big)$ (blue and red solid lines in Fig.~\ref{fig:1}a). This interrelation shows up in that $\omega_1(t)\omega_2(t)\!=\!\sqrt{\epsilon_{\mm}}\omega_{\mm}\omega_0$ or, alternatively---from the mentioned transcendental equation $\omega_l(t)$=$\sqrt{\frac{\epsilon_{\mm}}{\epsilon_l(t)}}\omega_{\mm}$, with $(l\!=\!1,2)$ and $\epsilon_l\!\overset{\scriptscriptstyle{\Delta}}{=}\!\epsilon_{\pp}(\omega_{l})$---, $\epsilon_1(t)\epsilon_2(t)\!=\!\epsilon_{\mm}(\frac{\omega_{\mm}}{\omega_0})^2$, allowing us to further write $\omega_1(t)\!=\!\sqrt{\epsilon_2(t)}\omega_0$ and $\omega_2(t)\!=\!\sqrt{\epsilon_1(t)}\omega_0$. Of course, by making $R\!\to\!\infty$, our original $\omega_{\mm}$ is instantaneously split into the final values of $(\omega_1,\omega_2)$, whereas making $R$ finite alters the dynamics of the problem: we have a transient and thus the amplitudes of the final forward and backward waves will be different. In addition, Fig.~\ref{fig:1}b shows the graphical match of momentum from the dispersion diagram of our Lorentzian when the blue solid line $\text{Re}(k_{\pp})$ crosses the dashed black line $k_{\mm}$. 

\section{Dynamics: Plane Wave(s) in a Time-Varying Lorentzian Medium}\label{sec:dynamics}

\subsection{Temporal Half-Space Scattering Coefficients}\label{subsec:dynamicsA}
In order to find the electromagnetic fields after the temporal discontinuity at $t\!=\!0$, we need to solve the wave equation subject to the temporal boundary conditions (BCs), including those stated at the end of Sec.~\ref{sec:Lorentzian}. One can find in the literature \cite{1124533,Xiao:14,8858032} that, in a nondispersive medium, it suffices to consider temporal continuity for both $D$ and $B$, which ensures that magnetic and electric fields $H$ and $E$ remain bounded, respectively: $D(z,t\!=\!0^{\pp})\!=\!D(z,t\!=\!0^{\mm})$ and $B(z,t\!=\!0^{\pp})\!=\!B(z,t\!=\!0^{\mm})$. This latter condition obviously becomes $H(z,t\!=\!0^{\pp})\!=\!H(z,t\!=\!0^{\mm})$ when magnetism is not present. In our case these two still apply, but two extra BCs are needed to determine the amplitudes of the forward and backward waves for both frequencies ($\omega_{1}$ and $\omega_{2}$): we can now use the fact---remarked in Sec.~\ref{sec:Lorentzian}---that $P(0^{\pp})\!=\!P(0^{\mm})$ and $\frac{dP(0^{\pp})}{dt}\!=\!\frac{P(0^{\mm})}{dt}$, where, e.g., $P(0^{\pm})$ stands for $P(z,t\!=\!0^{\pm})$ to reduce notation. Importantly, the joint continuities of $D$ and $P$ lead to the continuity of $E$: these three conditions are linearly dependent, so we choose to discard $P$. 

If we adopt the $e^{i\omega t}$ time-harmonic convention and use $k$=$k_{\mm}$=$k_{\pp}$, our initial forward waves can be written as (note that, in order to simplify notation, $E_{\mm}$ stands for $E(z,t\!<\!0)$, e.g.):
\begin{subequations}
\begin{gather}
E_{\mm}=e^{i\omega_{\mm}t}e^{-ikz},\label{eq:2B1a}\\
H_{\mm}=\frac{\sqrt{\epsilon_{\mm}}}{\eta_0}e^{i\omega_{\mm}t}e^{-ikz},\label{eq:2B1b}\\
D_{\mm}=\epsilon_0\epsilon_{\mm}e^{i\omega_{\mm}t}e^{-ikz},\label{eq:2B1c}\\
\frac{dP_{\mm}}{dt}=i\omega_{\mm}\epsilon_0(\epsilon_{\mm}-1)e^{i\omega_{\mm}t}e^{-ikz}.\label{eq:2B1d}
\end{gather}\label{eq:2B1}
\end{subequations}
Let us now see the complex space-time harmonic dependencies from a different perspective, and adopt the space-harmonic complex dependence $e^{-ikz}$, in which case forward and backward waves will be described by $e^{i\omega t}$ and $e^{-i\omega t}$, respectively. For $t\!>\!0$, the fields can be expressed as
\comment{
\begin{widetext}
\begin{subequations}
\begin{gather}
E_{\pp}=\left(f_1e^{i\omega_1t}+b_1e^{-i\omega_1t}+f_2e^{i\omega_2t}+b_2e^{-i\omega_2t}\right)e^{-ikz},\label{eq:2B2a}\\
H_{\pp}=\frac{1}{\eta_0}\left(\sqrt{\epsilon_1}(f_1e^{i\omega_1t}-b_1e^{-i\omega_1t})+\sqrt{\epsilon_2}(f_2e^{i\omega_2t}-b_2e^{-i\omega_2t})\right)e^{-ikz},\label{eq:2B2b}\\
D_{\pp}=\epsilon_0\left(\epsilon_1(f_1e^{i\omega_1t}+b_1e^{-i\omega_1t})+\epsilon_2(f_2e^{i\omega_2t}+b_2e^{-i\omega_2t})\right)e^{-ikz},\label{eq:2B2c}\\
\frac{dP_{\pp}}{dt}=i\epsilon_0\left(\omega_1(\epsilon_1-1)(f_1e^{i\omega_1t}-b_1e^{-i\omega_1t})+\omega_2(\epsilon_2-1)(f_2e^{i\omega_2t}-b_2e^{-i\omega_2t})\right)e^{-ikz}.\label{eq:2B2d}
\end{gather}\label{eq:2B2}%
\end{subequations}
\end{widetext}
}
\begin{subequations}
\begin{gather}
E_{\pp}=e^{-ikz}\sum_{l=1}^{2}\left(f_le^{i\omega_lt}+b_le^{-i\omega_lt}\right),\label{eq:2B2a}\\
H_{\pp}=e^{-ikz}\frac{1}{\eta_0}\sum_{l=1}^{2}\sqrt{\epsilon_l}\left(f_le^{i\omega_lt}-b_le^{-i\omega_lt}\right),\label{eq:2B2b}\\
D_{\pp}=e^{-ikz}\epsilon_0\sum_{l=1}^{2}\epsilon_l\left(f_le^{i\omega_lt}+b_le^{-i\omega_lt}\right),\label{eq:2B2c}\\
\frac{dP_{\pp}}{dt}=e^{-ikz}i\epsilon_0\sum_{l=1}^{2}\omega_l(\epsilon_l-1)\left(f_le^{i\omega_lt}-b_le^{-i\omega_lt}\right).\label{eq:2B2d}
\end{gather}\label{eq:2B2}%
\end{subequations}
where the unknowns $f_l$ and $b_l$ represent the amplitudes of the forward and backward electric field waves oscillating at frequency $\omega_l$. Enforcing time continuity of these four waves at $t\!=\!0$ leads---after some straightforward simplifications, replacing $\omega_l$=$\sqrt{\frac{\epsilon_{\mm}}{\epsilon_l}}\omega_{\mm}$, and using the BC for $H$ to simplify the BC for $\frac{dP}{dt}$---to the following system of equations:
\begin{equation}
    \begin{bmatrix}
        1 & 1 & 1 & 1 \\
        \sqrt{\epsilon_1} & -\sqrt{\epsilon_1} & \sqrt{\epsilon_2} & -\sqrt{\epsilon_2}  \\
        \epsilon_1 & \epsilon_1 & \epsilon_2 & \epsilon_2 \\
        \frac{1}{\sqrt{\epsilon_1}} & -\frac{1}{\sqrt{\epsilon_1}} & \frac{1}{\sqrt{\epsilon_2}} & -\frac{1}{\sqrt{\epsilon_2}}
    \end{bmatrix}
    \begin{bmatrix}
        f_1 \\
        b_1 \\
        f_2 \\
        b_2
    \end{bmatrix} 
    =
    \begin{bmatrix}
        1 \\
        \sqrt{\epsilon_{\mm}} \\
        \epsilon_{\mm} \\
        \frac{1}{\sqrt{\epsilon_{\mm}}}
    \end{bmatrix},\label{eq:2B3}  
\end{equation}
which gives us the closed-form solution to the unknown amplitudes:
\begin{subequations}
\begin{gather}
[f_1,b_1]=\frac{\epsilon_2-\epsilon_{\mm}}{2\sqrt{\epsilon_{\mm}}(\epsilon_2-\epsilon_1)}(\sqrt{\epsilon_{\mm}}\pm\sqrt{\epsilon_1}),\label{eq:2B4a}\\
[f_2,b_2]=\frac{\epsilon_{\mm}-\epsilon_1}{2\sqrt{\epsilon_{\mm}}(\epsilon_2-\epsilon_1)}(\sqrt{\epsilon_{\mm}}\pm\sqrt{\epsilon_2}),\label{eq:2B4b}
\end{gather}\label{eq:2B4}%
\end{subequations}
where $+$ ($-$) gives the forward $f_l$ (backward $b_l$) amplitudes. A set of analogous equations expressed only in terms of frequencies can be found in Appendix A.

In Fig.~\ref{fig:2}a we show the temporal evolution of the electromagnetic waves at $z\!=\!\lambda/16$ around the temporal jump (at $t\!=\!0$, indicated with black dashed lines) that results from Eqs.~(\ref{eq:2A1})-(\ref{eq:2B4}) when we consider the transition of Fig.~\ref{fig:1}  (the results obtained from FDTD simulations---marked with circles---when $N(t)$ follows the previously-mentioned $\text{tanh}(Rt)$ profile perfectly converge to these results as we make $R$ larger. Here we use $R\!=\!10^5/T$, with $T\!=\!\frac{2\pi}{\omega_{\mm}}$). 

\subsubsection{Approximations for $\omega_0\!\gg\!\omega_-$}\label{subsubsec:approx}
\begin{figure}[h]
\includegraphics[width=3.4in]{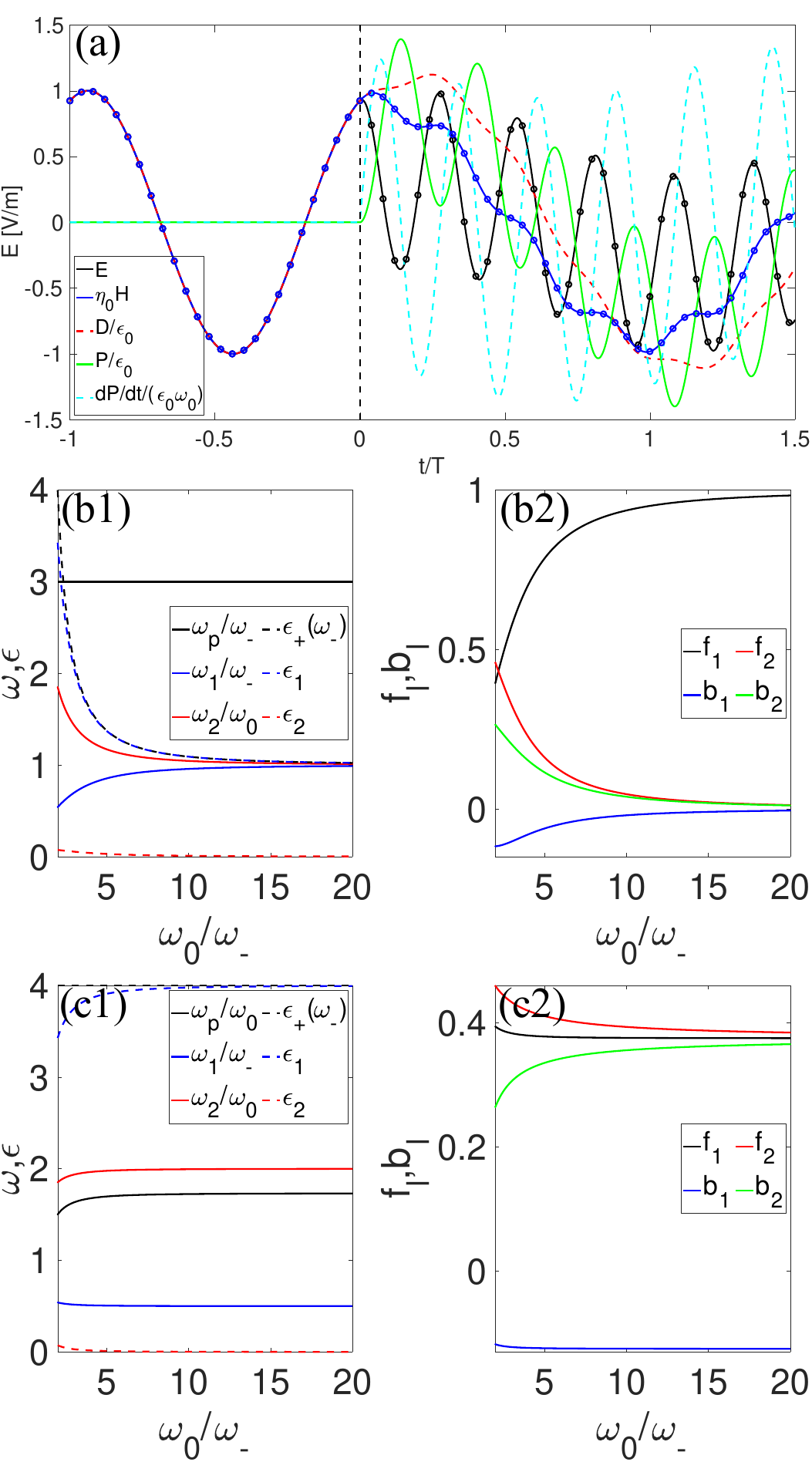}
\caption{(a) Electromagnetic waves vs. time at $z\!=\!\lambda/16$ for a transition from $\epsilon_{\protect\mm}(\omega_{\protect\mm})\!=\!1$ (vacuum) to $\epsilon_{\protect\pp}(\omega_{\protect\mm})\!=\!4$, with $\omega_0\!=\!2\omega_{\protect\mm}$ (the solid lines are analytical results, while the circular markers represent numerical FDTD simulations). (b1) New frequencies $\omega_i$ (and $\epsilon_{\protect\pp}(\omega_i)$) for $t\!>\!0$ and (b2) wave amplitude coefficients vs. $\omega_0/\omega_{\protect\mm}$, considering $\omega_{p\protect\pp}\!=\!3\omega_{\protect\mm}$ ($\epsilon_{\protect\pp}(\omega_{\protect\mm})\!\to\!1$ as $\omega_0/\omega_{\protect\mm}\!\to\!\infty$). Panels (c1)-(c2) are the same, but with $\epsilon_{\protect\pp}(\omega_{\protect\mm})\!=\!4$ ($\omega_{p\protect\pp}\!\to\!\sqrt{\chi_{\protect\pp}(\omega_{\protect\mm})}\omega_0$ as $\omega_0/\omega_{\protect\mm}\!\to\!\infty$).}
\label{fig:2}
\end{figure}

Now, let us ask ourselves what happens when $\omega_0$ increases, in which case we have to consider two different scenarios. In the (b) panels of Fig.~\ref{fig:2}, we keep $\omega_{p\pp}$ fixed with the value that makes $\epsilon_{\pp}(\omega_{\mm})\!=\!4$ when $\omega_0\!=\!2\omega_{\mm}$ and increase the ratio $\omega_0/\omega_{\mm}$: as this ratio tends to $\infty$, we have $\epsilon_1(\omega_1\!\to\!\omega_{\mm})\!\to\!1$ and $\epsilon_2(\omega_2\!\to\!\omega_0)\!\to\!0$ (panel (b1)), which makes $f_1\!\to\!1$ (see panel (b2)). Noting that $\sum_{l}\left(f_l+b_l\right)\!=\!0$, this means the initial plane wave is not altered by the temporal discontinuity, as one would expect from the fact that, given that the new medium is effectively transparent at $\omega_{\mm}$, no transfer of energy should take place between $\omega_{\mm}$ and $\omega_0$.

On the contrary, in the (c) panels we consider that $\omega_{p\pp}$ varies such that $\epsilon_{\pp}(\omega_{\mm})\!=\!4$ regardless of $\omega_0/\omega_{\mm}$ (see constant black dashed line in panel (c1)): by making $\omega_0/\omega_{\mm}\!\to\!\infty$, we now have $\omega_{p\pp}\!\to\!\infty$ and
\begin{subequations}
\begin{gather}
\epsilon_1\left(\omega_1\to\sqrt{\frac{\epsilon_{\mm}}{\epsilon_{\pp}(\omega_{\mm})}}\omega_{\mm}\right)\to\epsilon_{\pp}(\omega_{\mm}),\label{eq:2B5a}\\
\epsilon_2(\omega_2\!\to\!\sqrt{\epsilon_{\pp}(\omega_{\mm})}\omega_0)\!\to\!\frac{\epsilon_{\mm}}{\epsilon_{\pp}(\omega_{\mm})}\left(\frac{\omega_{\mm}}{\omega_0}\right)^2\to0,\label{eq:2B5b}
\end{gather}\label{eq:2B5}%
\end{subequations}
which transforms Eq.~\eqref{eq:2B4a} into $[f_1,b_1]\!=\!\frac{\sqrt{\epsilon_{\mm}}}{2\epsilon_1}(\sqrt{\epsilon_{\mm}}\pm\sqrt{\epsilon_1})$, i.e., the exact same expressions of the nondispersive scenario \cite{1124533,Xiao:14,8858032}. Nonetheless, we now also have $f_2\!=\!b_2\!=\!\frac{\epsilon_1-\epsilon_{\mm}}{2\epsilon_1}\!\neq\!0$, (in this precise example $f_2\!=\!b_2\!=\!f_1$, as depicted in panel (c2)): one thus has to wonder how to connect this solution including oscillations at $\omega_2\!\to\!\infty$ with the nondispersive situation, and first realize that the new medium is effectively $\epsilon$-near-zero (ENZ) \cite{PhysRevLett.97.157403,PhysRevLett.100.033903}. Substituting $\epsilon_2\!\to\!0$ into Eqs.~\eqref{eq:2B2} we see that $H_{\pp}(\omega_2)\!\to\!0$ (and $D_{\pp}(\omega_2)\!\to\!0$), making the Poynting vector at $\omega_2$ tend to zero and all of the power purely reactive. We must recognize, however, that as soon as we allow for some infinitesimally small loss (see Sec.~\ref{sec:lossy}), as required by our Lorentzian in order to become physical, $\omega_2$ becomes purely imaginary and its components immediately vanish (more details can be found in Appendix C). Noting that, when there is no dispersion, $E$ and $P$ are discontinuous across the temporal boundary---which entails a change of electromagnetic energy density---, our suddenly-vanishing $\omega_2$ components are nothing but the dispersive manifestation of this behavior.

\subsection{A First-Principles Approach: Use of Laplace Transform}\label{subsec:dynamicsB}
From Maxwell's equations with a general polarization vector,
\begin{subequations}
	\begin{align}
	\nabla\times\mathbf{E}&=-\mu_0\frac{\partial \mathbf{H}}{\partial t}\label{eq:p-eqa}\\
	\nabla\times\mathbf{H}&=\epsilon_0\frac{\partial\mathbf{E}}{\partial t}+\frac{\partial\mathbf{P}}{\partial t}+
	\mathbf{J}\label{eq:p-eqb}
	\end{align}
\end{subequations}
one can derive the pertinent wave equation
\begin{equation}\label{eq:p-eq_0}
\nabla\times	\nabla\times\mathbf{E}=-\mu_0\epsilon_0\frac{\partial^2\mathbf{E}}{\partial t^2}-\mu_0\frac{\partial^2\mathbf{P}}{\partial t^2}-\mu_0\frac{\partial 
	\mathbf{J}}{\partial t}
\end{equation}
Transforming into the Laplace domain, taking into account Eqs.~(\ref{eq:inital_P}),(\ref{eq:dpdt}) and restricting ourselves to $\omega_{p\mm}\!=\!0$, 
\begin{multline}\label{eq:p-eq_0_laplace_IC}
\frac{1}{\mu_0}\nabla\times	\nabla\times\tilde{\mathbf{E}}(\mathbf{r},s)=
\\
-\epsilon_0\left( s^2\tilde{\mathbf{E}}(\mathbf{r},s)-s\mathbf{E}(\mathbf{r},0^{\mm})-\frac{\partial\mathbf{E}(\mathbf{r},0^{\mm})}{\partial t}\right)
\\
-\left( s^2\tilde{\mathbf{P}}(\mathbf{r},s)+s\tilde{\mathbf{J}}(\mathbf{r},s)-\mathbf{J}(\mathbf{r},0^{\mm})\right).
\end{multline}

Now combine Eq.~\eqref{eq:p-eq_0_laplace_IC} with the constitutive relation Eq.~\eqref{eq:motiontime_4}, to obtain
\begin{multline}\label{eq:s-domain}
\frac{1}{\mu_0}\nabla\times	\nabla\times\tilde{\mathbf{E}}(\mathbf{r},s)+
\frac{\epsilon_0s^2(s^2+\omega_0^2+\omega_{p\pp}^2)}{s^2+\omega_0^2}\tilde{\mathbf{E}}(\mathbf{r},s)
\\=\epsilon_0\left( s\mathbf{E}(\mathbf{r},0^{\mm})+\frac{\partial\mathbf{E}(\mathbf{r},0^{\mm})}{\partial t}\right) 
-s\tilde{\mathbf{J}}(\mathbf{r},s)+\mathbf{J}(\mathbf{r},0^{\mm}).
\end{multline}

Let us take the one dimensional reduction of Eq.~\eqref{eq:p-eq_0} with $\mathbf{E}\!=\!\widehat{\mathbf{x}}E(z,t)$. In view of preservation of momentum we take $k\!=\!k_{\mm}\!=\!\omega_{\mm}\sqrt{\mu_0\epsilon_0}$ throughout. Also, $\nabla\!=\!-ik$, so from
\begin{equation}\label{eq:p-eq_51}
(k^2+\mu_0\epsilon_0\frac{\partial^2}{\partial t^2})E=-\mu_0\frac{\partial^2P}{\partial t^2}-\mu_0\frac{\partial J}{\partial t}
\end{equation}
Eq.~\eqref{eq:s-domain} becomes
\begin{multline}\label{eq:p-eq_0_laplac_11d1}
\epsilon_0\left( \omega_{\mm}^2
+
s^2\frac{s^2+\omega_0^2+\omega_{p\pp}^2}{s^2+\omega_0^2}\right) \tilde{E}(z,s)=
\\
\epsilon_0\left( sE(z,0^{\mm})+\frac{\partial E}{
	\partial t}(z,0^{\mm}) \right) 
-s\tilde{J}(z,s)+J(z,0^{\mm}),
\end{multline}
or
\begin{multline}\label{eq:p-eq_0_laplac_11d2}
\tilde{E}(z,s)=\\
(s^2+\omega_0^2)\frac{\epsilon_0\left( sE(z,\!0^{\mm})+\frac{\partial E}{
		\partial t}(z,\!0^{\mm}) \right) -s\tilde{J}(z,\!s)+J(z,\!0^{\mm})}{\epsilon_0(\omega_{\mm}^2s^2+\omega_{\mm}^2\omega_0^2+
	s^4+s^2\omega_0^2+s^2\omega_{p\pp}^2)}.
\end{multline}
The denominator of Eq.~\eqref{eq:p-eq_0_laplac_11d2} can be factored as
\begin{multline}\label{eq:p-eq_0_laplace_factored}
s^4+\left( \omega_{\mm}^2+\omega_0^2+\omega_{p\pp}^2\right)s^2 +\omega_{\mm}^2\omega_0^2=(s^2-s_1^2)(s^2-s_2^2)
\\
=(s^2+\omega_1^2)(s^2+\omega_2^2)
\end{multline}
with $s_l\!=\!\pm i\omega_l$. Note the agreement with the kinematic characteristic equation \eqref{eq:Lorentz-7-6}.
%

For $t\!<\!0$, the electric field is given as $E(z,t\!<\!0)\!=\!\cos{(\omega_{\mm}t-kz)}$. At the time $t\!=\!0^{\mm}$,
\begin{subequations}\label{eq:plane_wave_t_1}
	\begin{align}
E(z,t=0^{\mm})&=\cos{(kz)},\\
\frac{\partial E}{\partial t}(z,t=0^{\mm})&=\omega_{\mm}\sin{(kz)}.
\end{align}
\end{subequations}
We are now able to rewrite Eq.~\eqref{eq:p-eq_0_laplac_11d2} in the form
\begin{multline}\label{eq:p-eq_0_laplac_11d3}
\tilde{E}(z,s)
=
(s^2+\omega_0^2)\frac{sE(z,0^{\mm})+\frac{\partial E}{
		\partial t}(z,0^{\mm})}{(s^2+\omega_1^2)(s^2+\omega_2^2)}
\\
-
(s^2+\omega_0^2)\frac{s\tilde{J}(z,s)-J(z,0^{\mm})}{\epsilon_0(s^2+\omega_1^2)(s^2+\omega_2^2)}
\end{multline}
An inverse transform of Eq.~\eqref{eq:p-eq_0_laplac_11d3} for the source-free case yields
\begin{equation}\label{key}
E(z,t)=E_1^{\pp}+E_1^{\mm}+E_2^{\pp}+E_2^{\mm}
\end{equation}
where
\begin{subequations}\label{eq:e12}
\begin{align}
E_1^\pm=&\frac{ \omega_0^2-\omega_1^2}{\omega_2^2-\omega_1^2} \frac{1}{2}  \left(1\pm \frac{\omega_{\mm}}{\omega_1}\right)\cos{(\omega_1t\mp kz)},\\
E_2^\pm=&\frac{ \omega_0^2-\omega_2^2}{\omega_2^2-\omega_1^2} \frac{1}{2}  \left(1\pm \frac{\omega_{\mm}}{\omega_2}\right)\cos{(\omega_2t\mp kz)},
\end{align}
\end{subequations}
which are the same exact expressions that result from keeping the real part of Eq.~(\ref{eq:2B2a}), with $[f_l,b_l]$ from Eqs.~(\ref{eq:2B4}) (or, more directly, Eqs.~(\ref{eq:A2})) simplified through $\epsilon_{\mm}\!=\!1$. Under the approximations of Eqs.~(\ref{eq:2B5}), the latter results simplify to
\begin{subequations}\label{eq:e12pm}
	\begin{align}
	E_1^\pm&\simeq\frac{ 1\pm\sqrt{\epsilon_2}}{2\epsilon_2}\cos{(\omega_1t\mp kz)}\\
	E_2^\pm&\simeq\frac{ \epsilon_2-1}{2\epsilon_2} \left(1\pm\frac{1}{\frac{\omega_0}{\omega_{\mm}}\sqrt{\epsilon_2}}\right)\cos{(\omega_2t\mp kz)}.
	\end{align}
\end{subequations}
A corresponding approximation for the magnetic field is then
\begin{subequations}\label{eq:h_field_1}
	\begin{align}
	H_1^\pm&\simeq\pm\frac{ 1\pm \sqrt{\epsilon_2}}{2\sqrt{\epsilon_2}}\frac{1}{\eta_0}\cos{(\omega_1t\mp kz)}\\
	H_2^\pm&\simeq\nonumber
	\\\pm\frac{ \epsilon_2-1}{2\epsilon_2}& \left(1\pm\frac{\omega_{\mm}}{\omega_0\sqrt{\epsilon_2}}\right)\frac{\omega_{\mm}}{ \omega_0\sqrt{\epsilon_2}}\frac{1}{\eta_0}\cos{(\omega_2t\mp kz)}\approx 0.
	\end{align}
\end{subequations}

\section{A Transmission-Line Model}\label{sec:transmission_line}
The time-varying Lorentzian response described in Eq.~\eqref{eq:202} can be viewed as the (polarization) charge response to an applied voltage across a series time-varying LC circuit, and thus rewritten as
\begin{equation}
L(t)\frac{d^2P(t)}{dt^2}+\frac{1}{C(t)}P(t)=E(t),\label{eq:401}
\end{equation}
with $L(t)\!=\!\frac{1}{\epsilon_0\omega_p^2(t)}$ and $C$=$\frac{1}{\omega_0^2L(t)}$=$\epsilon_0\left(\frac{\omega_p(t)}{\omega_0}\right)^2$. Two important facts must be pointed out here: (i) $\omega_0$ is kept constant and (ii) there is no $\frac{dL(t)}{dt}\frac{dP(t)}{dt}$ term. Accordingly, considering that, for $t\!<\!0$, our dispersive medium is modeled as a transmission line with an $L_{\mm}C_{\mm}$ branch, we can think of our sudden change in $\omega_p(t)$ as the connection of a new $L_pC_p$ branch in parallel, with $L_p$=$\frac{L_{\mm}L_{\pp}}{L_{\mm}-L_{\pp}}$ and $C_p$=$\frac{1}{\omega_0^2L_p}$=$C_{\pp}\!-\!C_{\mm}$ (note that both $L_p$ and $C_p$ will be non-Foster when $\omega_{p\pp}\!<\!\omega_{p\mm}$, and note also that disconnecting the $L_{\mm}C_{\mm}$ branch will turn the medium into vacuum. These aspects will be discussed in our future study); when $\omega_{p\mm}\!=\!0$ (vacuum), $L_{\mm}\!=\!\infty$ and $C_{\mm}\!=\!0$, and thus $L_p\!=\!L_{\pp}$ and $C_p\!=\!C_{\pp}$. Now, the inductor $L_p$ forbids a discontinuity in $P_p$ that would generate a spike of polarization current $\frac{dP_p}{dt}$ across the new branch: $P_p(0^{\pp})\!=\!P_p(0^{\mm})\!=0$ ($v_{C_p}(0^{\pp})\!=\!v_{C_p}(0^{\mm})\!=\!0$). Besides, there can be no discontinuity in the magnetic flux linkage $\Phi_p\!=\!L_p\frac{dP_p}{dt}$---the fact that we use the term ``magnetic'' should not give rise to confusion: we are using inductors to model the dispersive behavior of the dielectric function but there is no magnetism involved; more specifically, in the picture of a mass-spring oscillator, the inductor represents mass and is therefore related to (mechanical) momentum and kinetic energy, whereas the capacitor models the spring constant and is related to potential energy. A further argument is that, unlike $\Phi_{L_0}$, $\Phi_{L_{\mm}}$ and $\Phi_{L_p}$ are related to $\frac{dH}{dz}$, not to $H$, and hence $v_{L_{\mm}}$ and $v_{L_p}$ are related to $E$, not to $\frac{dE}{dz}$---that would lead to a spike in $v_{L_p}$: $\frac{dP_p(0^{\pp})}{dt}\!=\!\frac{dP_p(0^{\mm})}{dt}\!=0$. Therefore, we finally have $\frac{d^2P_p(0^{\pp})}{dt^2}\!=\!\frac{E(0^{\pp})}{L_p}\!=\frac{E(0^{\mm})}{L_p}$. Towards the end of Sec.~\ref{sec:Lorentzian}, continuity conditions for $P$ and $\frac{dP}{dt}$ were derived from a functional-analysis point of view of our LTI system's response; as it is clear that the voltages and currents across the $L_{\mm}C_{\mm}$ branch are also continuous, we have now arrived, from a circuital perspective, to the same continuity conditions. 
\begin{figure}[h]
\includegraphics[width=3.4in]{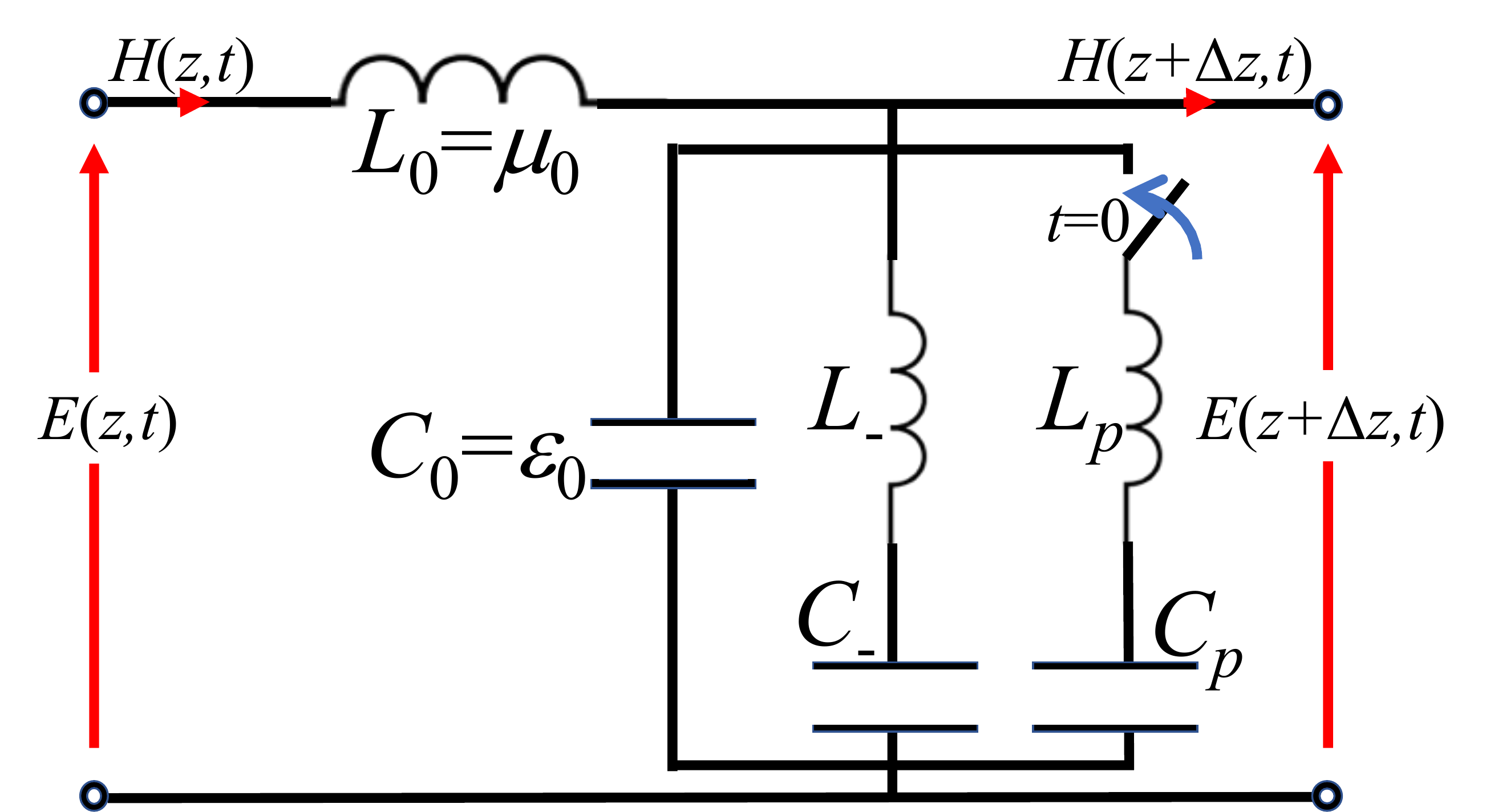}
\caption{Transmission-line equivalence of our unbounded time-varying dispersive medium. At $t\!=\!0$, the switch is closed, effectively connecting our series tank circuit $L_pC_p$.}
\label{fig:3}
\end{figure}

\section{The Lossy Case}\label{sec:lossy}
\begin{figure}[h!]
\includegraphics[width=3.4in]{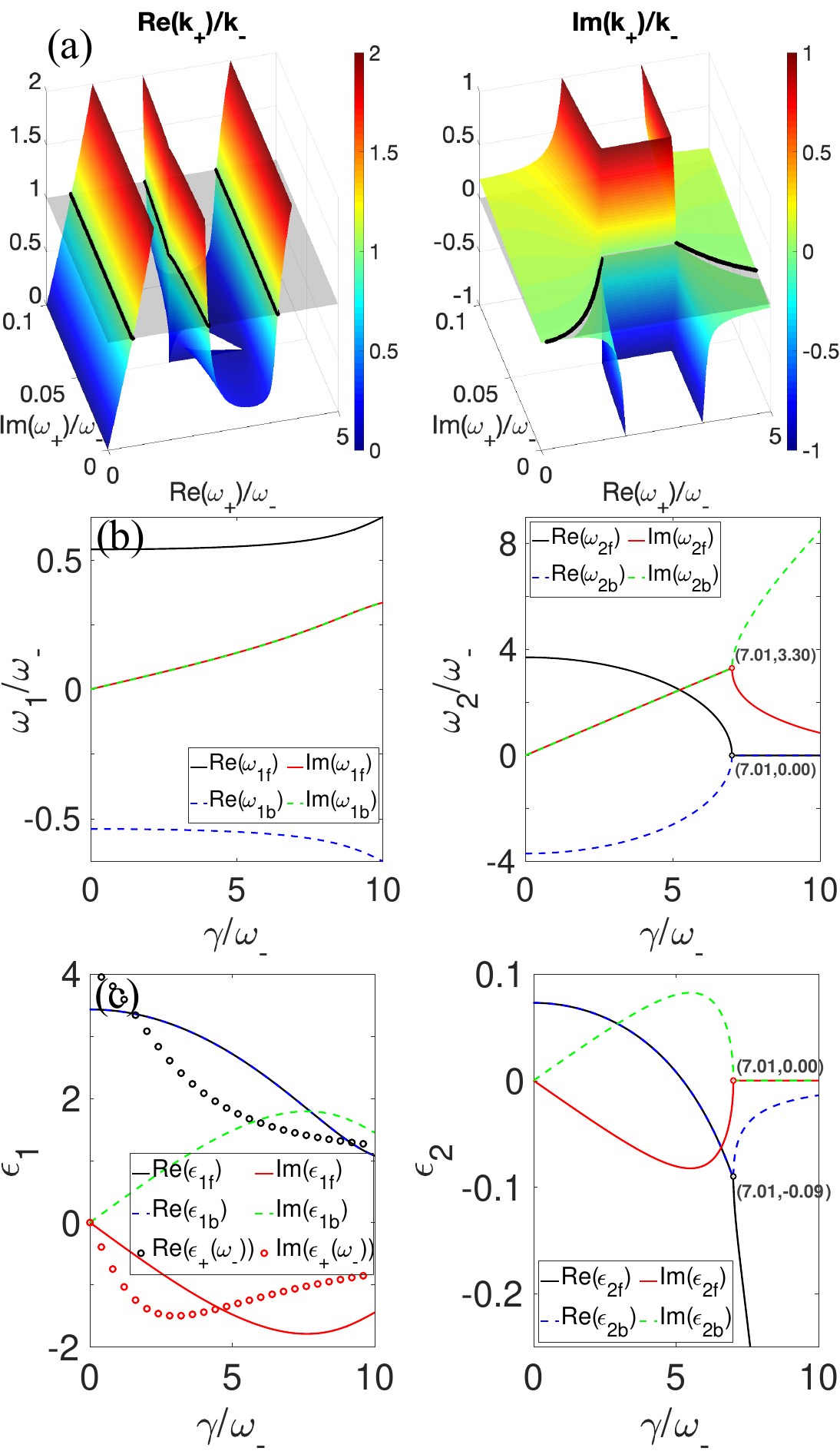}
\caption{(a) Dispersion diagram in the complex $k$- and $\omega$-planes when $\epsilon_{\protect\pp}(\omega_{\protect\mm})\!=\!4\!-\!0.1i$, which results in $\omega_{p\protect\pp}\!\approx\!3\omega_{\protect\mm}$ and $\gamma\!=\!0.1\omega_{\protect\mm}$ when $\omega_0\!=\!2\omega_{\protect\mm}$. Conservation of momentum is achieved when the surfaces $[\omega_{\protect\pp},\text{Re}\left(k_{\protect\pp}(\omega_{\protect\pp})\right)/k_{\protect\mm}]$ and $[\omega_{\protect\pp},\text{Im}\left(k_{\protect\pp}(\omega_{\protect\pp})\right)/k_{\protect\mm}]$ simultaneously intersect the $\text{Re}(k_{\protect\pp})/k_{\protect\mm}\!=\!1$ and $\text{Im}(k_{\protect\pp})/k_{\protect\mm}\!=\!0$ planes (grey color), respectively. These intersection curves are marked in black, and give rise to four complex frequencies that, in this example, form two complex-conjugate pairs in the $s$-plane (the plotted region $\text{Re}(\omega_{\protect\pp})\!>\!0$ only includes one complex frequency per pair). (b) Evolution of the four complex frequencies vs. $\gamma$, with the other parameters fixed ($\omega_0$ and $\omega_{p\protect\pp}$ from panel (a)): at $\gamma\!=\!7.01\omega_{\protect\mm}$ we reach a critical point and the pair of complex frequencies linked to $\omega_0$ ($\omega_{2f}$ and $\omega_{2b}$) splits into two purely imaginary frequencies for which $\epsilon$ becomes purely real and negative; the latter is seen in panel (c). Note that, in this overdamped region ($\gamma\!>\!7.01\omega_{\protect\mm}$), the notation $\omega_{2f}$ and $\omega_{2b}$ is not strictly rigorous: this pair simply becomes $\omega_2$ and $\omega_3$, and the associated waves represent non-propagating damping.}
\label{fig:4}
\end{figure}

\begin{figure}[h]
\includegraphics[width=3.4in]{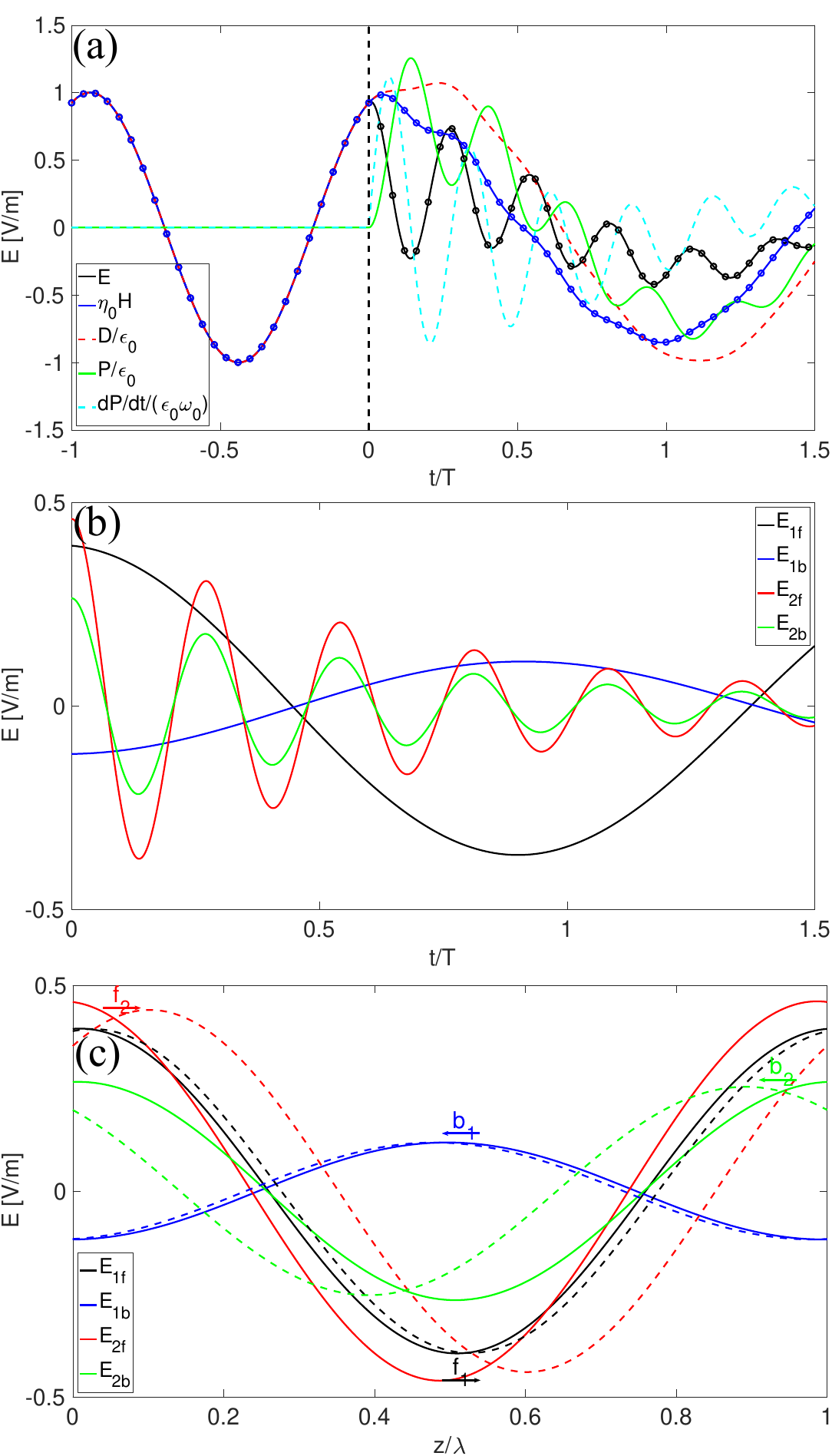}
\caption{(a) Electromagnetic waves vs. time at $z\!=\!\lambda/16$ when $\omega_0$ and $\omega_{p\protect\pp}$ are taken from Fig.~\ref{fig:4} and  $\gamma\!=\!0.5\omega_{\protect\mm}$ (numerical FDTD simulations are marked with circles): we have $\omega_{2f}$ and $\omega_{2b}$, corresponding with an underdamped scenario. (b) Separate components of $E(z\!=\!0,t\!>\!0)$. (c) Two snapshots of the separate components of $E_{\protect\pp}$ (the solid and dashed lines represent $E(z,t\!=\!0^{\protect\pp})$ and $E(z,t\!=\!T/32)$, respectively), showing forward and backward propagation for both $\omega_1$ and $\omega_2$.}
\label{fig:5}
\end{figure}

If we introduce loss into our time-varying Lorentzian medium, Eqs.~(\ref{eq:202}) and (\ref{eq:401}) must be extended as
\begin{subequations}
\begin{gather}
\frac{d^2P(t)}{dt^2}+\gamma\frac{dP(t)}{dt}+\omega_0^2P(t)=\epsilon_0\omega_p^2(t)E(t),\label{eq:501a}\\
L(t)\frac{d^2P(t)}{dt^2}+R(t)\frac{dP(t)}{dt}+\frac{1}{C(t)}P(t)=E(t),\label{eq:501b}
\end{gather}\label{eq:501}%
\end{subequations}
with $R(t)\!=\!\gamma L(t)$. For conciseness, we will not write down here the lengthy expressions of the complex frequencies that enforce $k_{\pp}\!=\!k_{\mm}$ (panel (a) in Fig.~\ref{fig:4} shows the complex-frequency dispersion diagram for  $\epsilon_{\pp}(\omega_{\protect\mm})\!=\!4\!-\!0.1i$ with $\omega_0\!=\!2\omega_{\protect\mm}$) according to
\begin{equation}
\omega_{\mm}\sqrt{1+\frac{\omega_{p\mm}^2}{\omega_0^2-\omega_{\mm}^2+i\omega_{\mm}\gamma}}=\omega_{\pp}\sqrt{1+\frac{\omega_{p\pp}^2}{\omega_0^2-\omega_{\pp}^2+i\omega_{\pp}\gamma}},\label{eq:502}
\end{equation}
but it is worth pointing out that three different scenarios open up. We will now restrict the discussion to the particular case where $\omega_{p\mm}\!=\!0$ (further insights will be presented in an upcoming study). If, starting from $\gamma\!=\!0$, we gradually increase loss, a positive imaginary part---note that, given that we are adopting the $e^{i\omega t}$ convention, $\text{Im}(\omega_{\pp})\!>\!0$ represents frequencies that are damped---begins to show up in the two pairs of solutions from Eq.~\eqref{eq:2A2} (this is seen in panels (b) of Fig.~\ref{fig:4}, depicting the variation of these complex frequencies with $\gamma/\omega_{\mm}$), so we have two distinct pairs of complex frequencies (complex conjugate pairs in the Laplace transform $s$-plane): $\pm\omega_{lr}\!+\!i\omega_{li}$, with $\omega_{lr}$ and $\omega_{li}$ real and positive ($\omega_{lf}\!=\!+\omega_{lr}\!+\!i\omega_{li}$ and $\omega_{lb}\!=\!-\omega_{lr}\!+\!i\omega_{li}$ will therefore describe forward- and backward-propagating evanescent waves, respectively). Each pair can then be seen as the two characteristic roots of the natural response of some underdamped RLC oscillator, and the forward and backward waves for frequency $l$ will have the form $e^{-\omega_{li}t}\text{cos}(\omega_{lr} t\pm kz+\varphi)$, $\varphi$ being a phase term. If we define the $s$-plane frequencies $s_l\!=\!i\omega_{lr}\!-\!\omega_{li}$, the electromagnetic waves for $t\!>\!0$ can be described as
\begin{subequations}
\begin{gather}
E_+=e^{-ikz}\sum_{l=1}^{2}\left(f_le^{s_lt}+b_le^{s_l^*t}\right),\label{eq:503a}\\
H_+=e^{-ikz}\frac{1}{\eta_0}\sum_{l=1}^{2}\left(\sqrt{\epsilon_l}f_le^{s_lt}-\sqrt{\epsilon_l^*}b_le^{s_l^*t}\right),\label{eq:503b}\\
D_+=e^{-ikz}\epsilon_0\sum_{l=1}^{2}\left(\epsilon_lf_le^{s_lt}+\epsilon_l^*b_le^{s_l^*t}\right),\label{eq:503c}\\
\frac{dP_+}{dt}=e^{-ikz}\epsilon_0\sum_{l=1}^{2}\left(s_l\chi_lf_le^{s_lt}+s_l^*\chi_l^*b_le^{s_l^*t}\right),\label{eq:503d}
\end{gather}\label{eq:503}%
\end{subequations}
and thereby the unknown amplitudes can be calculated as
\begin{equation}
    \begin{bmatrix}
        1 & 1 & 1 & 1 \\
        \sqrt{\epsilon_1} & -\sqrt{\epsilon_1^*} & \sqrt{\epsilon_2} & -\sqrt{\epsilon_2^*}  \\
        \epsilon_1 & \epsilon_1^* & \epsilon_2 & \epsilon_2^* \\
        s_1\chi_1 & s_1^*\chi_1^* & s_2\chi_2 & s_2^*\chi_2^*
    \end{bmatrix}
    \begin{bmatrix}
        f_1 \\
        b_1 \\
        f_2 \\
        b_2
    \end{bmatrix} 
    =
    \begin{bmatrix}
        1 \\
        \sqrt{\epsilon_{{\mm}}} \\
        \epsilon_{\mm} \\
        s_{\mm}\chi{\mm}
    \end{bmatrix},\label{eq:504}  
\end{equation}
where $s_{\mm}\!=\!i\omega_{\mm}$. This character of the waves, decaying with $t$ but not with $z$, is clearly seen in panels (b) and (c) of Fig.~\ref{fig:5}, which depicts the underdamped scenario associated with $\gamma\!=\!0.5\omega_{\mm}$.

\begin{figure}[h]
\includegraphics[width=3.4in]{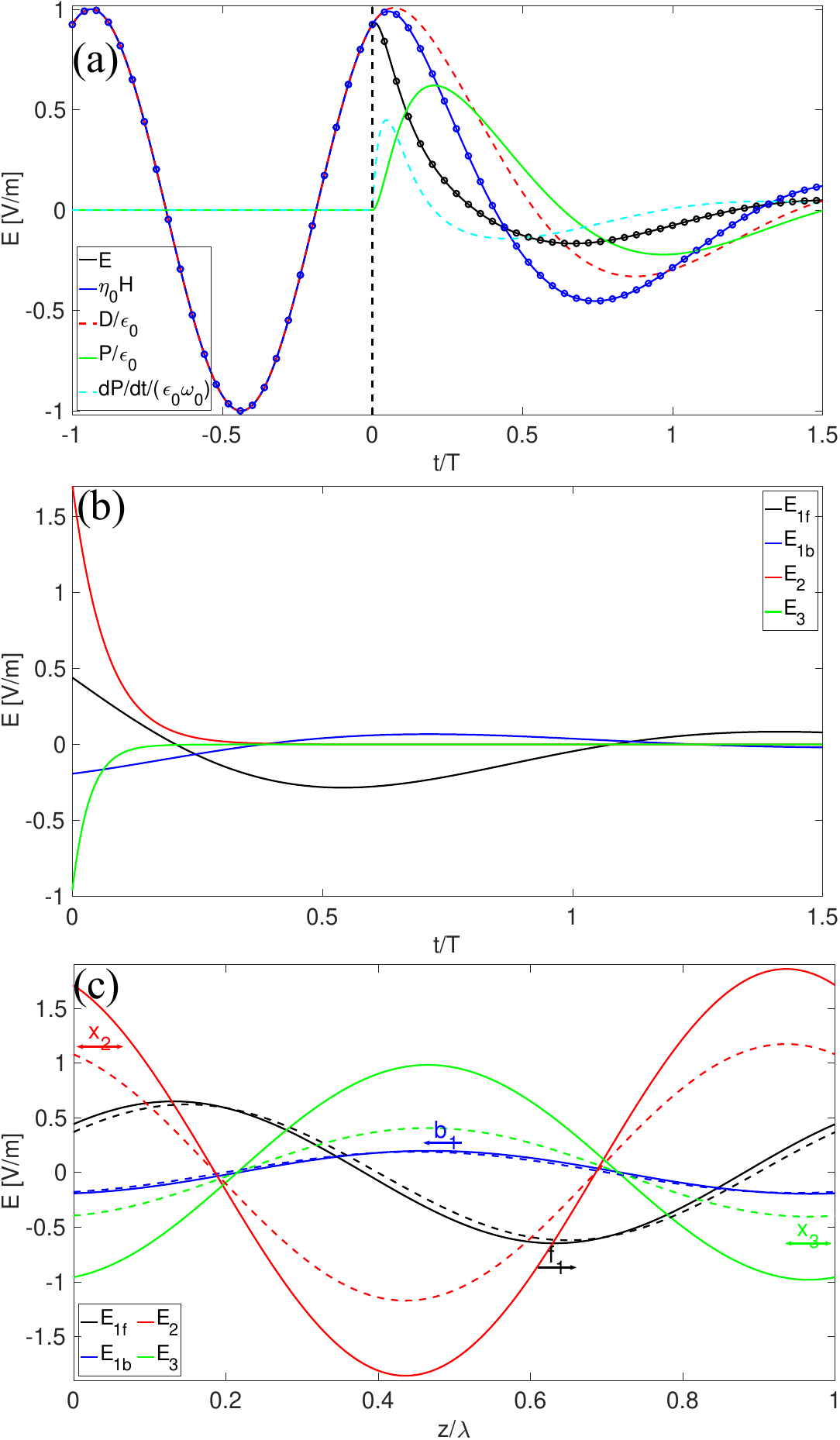}
\caption{Same as Fig.~\ref{fig:5} but with $\gamma\!=\!7.3\omega_{\protect\mm}$, yielding an overdamped regime with purely imaginary $\omega_2$ and $\omega_3$ describing oscillations that do not propagate. This is seen in panel (c): red ($E_{2}$) and green ($E_{3}$) curves.}
\label{fig:6}
\end{figure}

\begin{figure}[h]
\includegraphics[width=3.4in]{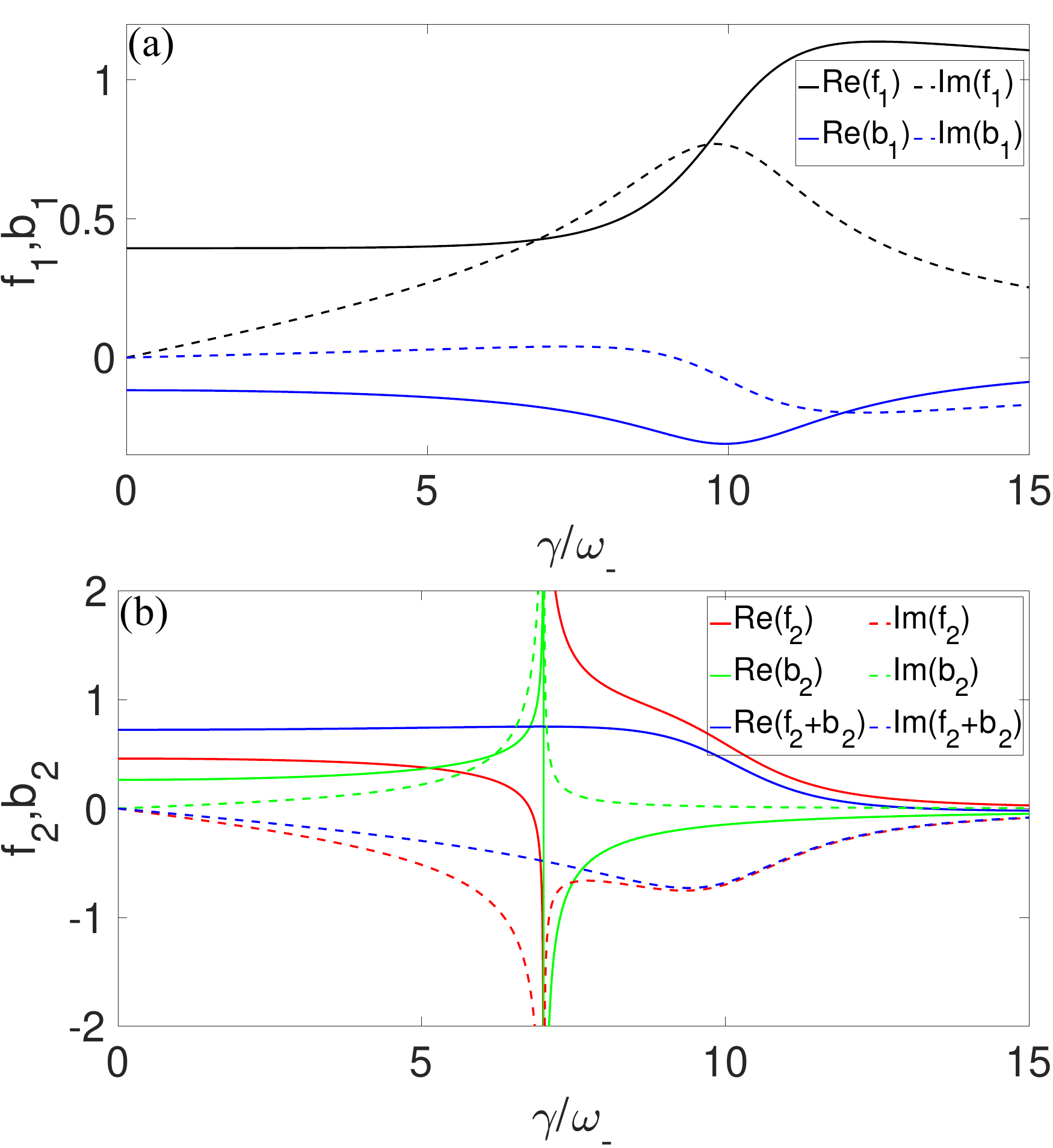}
\caption{Real and imaginary part of the complex amplitude coefficients vs. $\gamma/\omega_{\protect\mm}$. As $\omega_{2f}\!\to\!\omega_{2b}$ near the point of critical damping $\gamma\!=\!7.01\omega_{\protect\mm}$, $f_2$ and $b_2$ tend to diverge but with opposite signs (panel (b)), keeping $f_2\!+\!b_2$ (or else $x_2\!+\!x_3$) bounded. }
\label{fig:7}
\end{figure}

If we keep increasing $\gamma$, we will reach a critical point ($\gamma\!=\!7.01\omega_{\mm}$ in Fig.~\ref{fig:4}) at which the second pair of complex frequencies collapses into the same purely imaginary frequency $+i\omega_{2i}$, so one can think of this pair as the two equal characteristic roots of some critically damped RLC oscillator. Propagation for $+i\omega_{2i}$ is obviously forbidden, with $\epsilon_2$ purely real and negative (see Fig.~\ref{fig:4}, panels (c)), and the waves will have the form $e^{-\omega_{2i}t}\text{cos}(kz+\varphi)$. Also, assuming $\omega_{\mm}\!<\!\omega_0$, in general we have ${\lvert}\epsilon_2{\rvert}\ll1$. Further, if $\gamma$ is increased beyond the point of critical damping, $+i\omega_{2i}$ is split into two different purely imaginary frequencies, as corresponds to an overdamped RLC oscillator, which we will denote $\omega_2$ and $\omega_3$ (the retrieval of the temporal half-space scattering coefficients is described in Appendix B). This time-decaying non-propagating nature associated with $\omega_2$ and $\omega_3$ is illustrated in the overdamped scenario of Fig.~\ref{fig:6} ($\gamma\!=\!7.3\omega_{\mm}$), see red and green plots in panels (b) and (c). Finally, Fig.~\ref{fig:7} depicts the evolution of the scattering coefficients with $\gamma/\omega_{\mm}$ and how $f_2\!+\!b_2$ ($x_2\!+\!x_3$ after the critical point) remains bounded, despite these coefficients separately diverging.

Incidentally, only when $\omega_{p\mm}\!=\!0$ do we have $s_1$ and $s_1^*$ (and $s_2$ and $s_2^*$ in the underdamped case). In general, for $\omega_{p\mm}\!<\!\omega_{p\pp}$, the characteristic roots $s_{1f}$ and $s_{1b}$ will approximately, but not exactly, form a complex conjugate pair. As a consequence, $\epsilon_{\pp}^*(-is_{1f})\neq\epsilon_{\pp}(-is_{1b})$, meaning that forward and backward waves do not propagate in the very same medium.

\section{Conclusions}\label{sec:conclusions}
We investigate the ``reflection/transmission'' of a monochromatic plane wave at a dispersive temporal boundary, substantiated as a step-like change in the plasma frequency of a Lorentz-type dielectric function, and we present a transmission-line equivalent modelling this transition. The fact that two frequencies rather than one, each with forward and backward propagating constituents, are instantaneously generated after the transition is in line with the second-order nature of the dispersion in the medium. When we omit loss, we can still connect this behavior with the well-known dispersionless case and show how, as $\omega_0/\omega_{\mm}$ increases, the lower frequency $\omega_1$ tends to the dispersionless solution, whereas the upper frequency $\omega_2$, linked to $\omega_0$, presents a markedly different phenomenon: not only does the medium acquire ENZ character at $\omega_2$, but also the forward and backward waves' amplitudes tend to converge, effectively constituting a standing wave along $z$ which, in the limit of negligible loss, almost instantaneously fades out. Importantly, one can see from the mathematics developed that the described analogy, exemplified in this work for a transition from free-space, also holds for the inverse transition to free-space, or any other transition for that matter. In an upcoming study, the issue of power storage/conveyance and conversion will be addressed in depth, but it is already evident from the above discussion that, in the $\omega_0/\omega_{\mm}\!\to\!\infty$ limit, no power propagates at $\omega_2$. 

\begin{acknowledgments}
This work is supported in part by the Vannevar Bush Faculty Fellowship program sponsored by the Basic Research Office of the Assistant Secretary of Defense for Research and Engineering, funded by the Office of Naval Research through grant N00014-16-1-2029. 
\end{acknowledgments}

\appendix                                     
\section{Scattering Coefficients for the Lossless Scenario in Terms of $\omega$}
We can substitute $\sqrt{\epsilon_l}$=$\frac{\omega_{\mm}}{\omega_l}\sqrt{\epsilon_{\mm}}$ in Eq.~\eqref{eq:2B3} to arrive at a set of equations expressed only in terms of frequencies:
\begin{equation}
    \begin{bmatrix}
        1 & 1 & 1 & 1 \\
        \frac{\omega_{\mm}}{\omega_1} & -\frac{\omega_{\mm}}{\omega_1} & \frac{\omega_{\mm}}{\omega_2} & -\frac{\omega_{\mm}}{\omega_2}  \\
        (\frac{\omega_{\mm}}{\omega_1})^2 & (\frac{\omega_{\mm}}{\omega_1})^2 & (\frac{\omega_{\mm}}{\omega_2})^2 & (\frac{\omega_{\mm}}{\omega_2})^2 \\
        \frac{\omega_1}{\omega_{\mm}} & -\frac{\omega_1}{\omega_{\mm}} & \frac{\omega_2}{\omega_{\mm}} & -\frac{\omega_2}{\omega_{\mm}}
    \end{bmatrix}
    \begin{bmatrix}
        f_1 \\
        b_1 \\
        f_2 \\
        b_2
    \end{bmatrix} 
    =
    \begin{bmatrix}
        1 \\
        1 \\
        1 \\
        1
    \end{bmatrix}.\label{eq:A1}  
\end{equation}
Note that, as the elements of the right hand side are all equal, this system is perfectly conditioned for numerical solving. The expressions for the unknown amplitudes in Eq.~\eqref{eq:2B4} thus have the alternative form:
\begin{subequations}
\begin{gather}
[f_1,b_1]=\frac{1}{2}\frac{\omega_2^2-\omega_{\mm}^2}{\omega_2^2-\omega_1^2}\frac{\omega_1}{\omega_{\mm}^2}(\omega_1\pm\omega_{\mm}),\label{eq:A2a}\\
[f_2,b_2]=\frac{1}{2}\frac{\omega_{\mm}^2-\omega_1^2}{\omega_2^2-\omega_1^2}\frac{\omega_2}{\omega_{\mm}^2}(\omega_2\pm\omega_{\mm}).\label{eq:A2b}
\end{gather}\label{eq:A2}%
\end{subequations}

\section{Scattering Coefficients in a Lossy Overdamped Scenario}
Given that we now have purely imaginary $\omega_2$ and $\omega_3$, which describe no propagation, the coefficients $f_2$ and $b_2$ are replaced with $x_2$ and $x_3$. The matrix system of equations becomes
\begin{equation}
    \begin{bmatrix}
        1 & 1 & 1 & 1 \\
        \sqrt{\epsilon_1} & -\sqrt{\epsilon_1^*} & \sqrt{\epsilon_2} & \sqrt{\epsilon_3}  \\
        \epsilon_1 & \epsilon_1^* & \epsilon_2 & \epsilon_3 \\
        s_1\chi_1 & s_1^*\chi_1^* & s_2\chi_2 & s_3\chi_3
    \end{bmatrix}
    \begin{bmatrix}
        f_1 \\
        b_1 \\
        x_2 \\
        x_3
    \end{bmatrix} 
    =
    \begin{bmatrix}
        1 \\
        \sqrt{\epsilon_{{\mm}}} \\
        \epsilon_{\mm} \\
        s_{\mm}\chi{\mm}
    \end{bmatrix}.\label{eq:B1}  
\end{equation}

\section{Adding a Small Loss when $\omega_0\!\to\!\infty$}
\renewcommand\thefigure{\thesection.\arabic{figure}} 
\setcounter{figure}{0}
We saw in Sec.~\ref{subsec:dynamicsA} in the main text (see panels (c1),(c2) of Fig.~\ref{fig:2}) how, for a given prescribed value of (lossless) $\epsilon_{\pp}(\omega_{\mm})$, taking the limit $\omega_0/\omega_{\mm}\!\to\!\infty$ leads to a situation that is equivalent to the well-known problem of a temporal half-space in a nondispersive medium, except for the fact that we now have additional forward and backward oscillations at $\omega_2\!\to\!\infty$---for which the medium is ENZ ($\epsilon_2\!\to\!0$)---with nonzero amplitudes $f_2\!=\!b_2\!=\!\frac{\epsilon_1-\epsilon_{\mm}}{2\epsilon_1}$. We also stated how adding an infinitesimally small amount of loss would lead  to instantaneously-vanishing $\omega_2$ components, thereby drawing an exact correspondence with the nondispersive scenario. Let us see this behaviour in more detail with the numerical example of Fig.~\ref{fig:C1}.
\begin{figure}[h]
\includegraphics[width=3.4in]{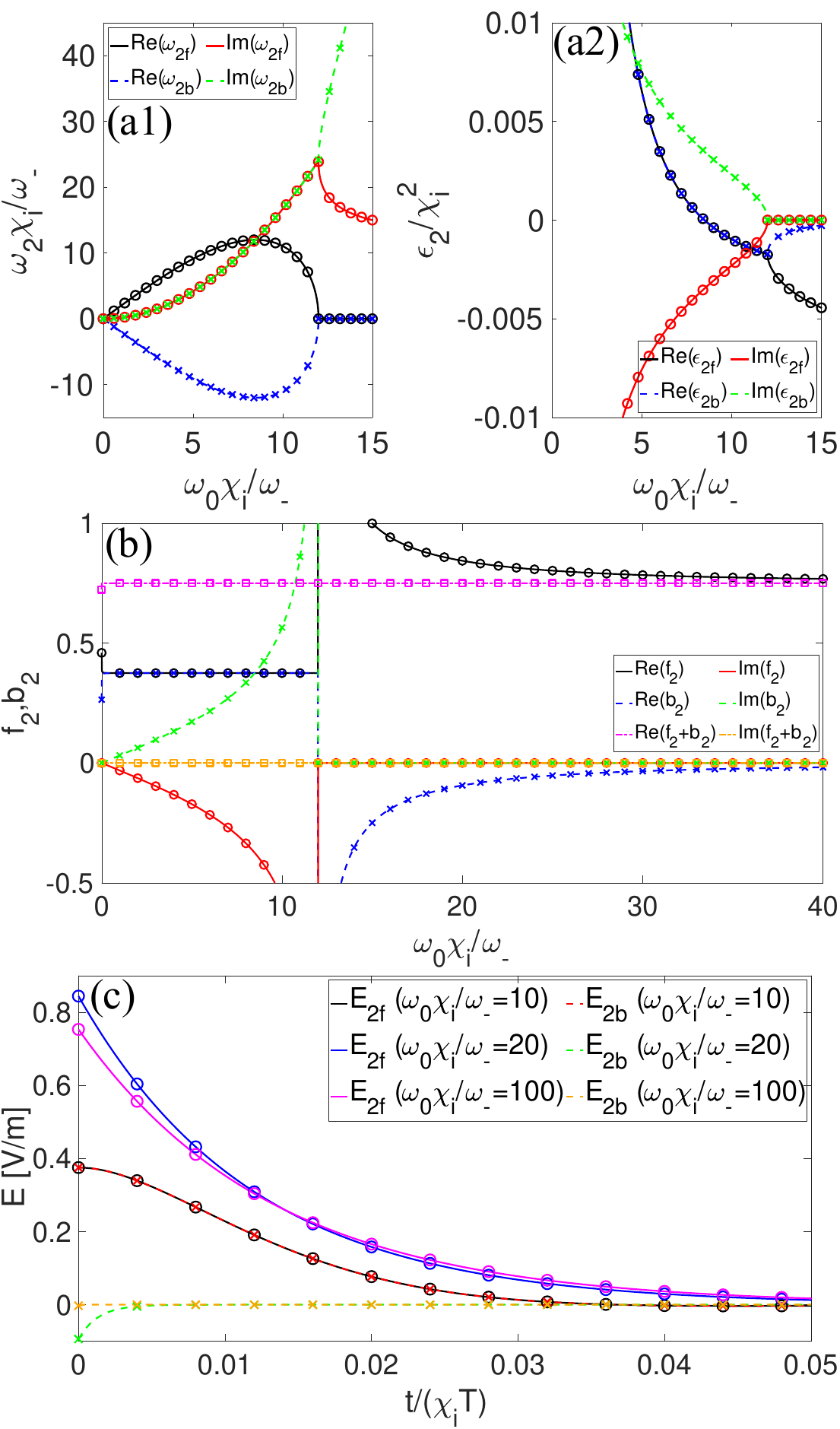}
\caption{(a1) Normalized real and imaginary part of the complex frequencies $\omega_{2f}$ ($\omega_2$) and $\omega_{2b}$ ($\omega_3$) vs. $\omega_0\chi_i/\omega_{\protect\mm}$, for $\chi_i\!=\!10^{-3}$ (lines) and $\chi_i\!=\!10^{-4}$ (markers). (a2) Normalized dielectric functions $\epsilon_{2f}$ ($\epsilon_2$) and $\epsilon_{2b}$ ($\epsilon_3$). (b) Complex amplitude coefficients $f_2$ ($x_2$) and $b_2$ ($x_3$). (c) $E_{2f}$ ($E_3$) and $E_{2b}$ ($E_3$) vs. normalized time, for $z\!=\!0$ and several $\omega_0/\omega_{\protect\mm}$ ratios; note how the black and red plots, corresponding to the underdamped region where $f_2\!=\!b_2$, are superimposed.}
\label{fig:C1}
\end{figure}

Panel (a1) shows the real and imaginary parts of the complex frequencies $\omega_{2f}$ and $\omega_{2b}$, which form a complex-conjugate pair in the $s$-plane when $\omega_0/\omega_{\mm}$ is smaller than the point of critical damping (see Sec.~\ref{sec:lossy}). Beyond the critical point, this pair becomes purely imaginary: denoting $\chi_{\pp}(\omega_{\mm})$ by $\chi_r\!-\!i\chi_i$, it can be shown that, as $\omega_0/\omega_{\mm}\!\to\!\infty$, we have $\gamma\!\to\!\frac{\chi_i}{\chi_r}\frac{\omega_0^2}{\omega_{\mm}}$ and $\omega_3\!\to\!i\gamma\!\to\!i\infty$, together with $\omega_2\!\to\!i\frac{(\chi_r+1)\chi_r}{\chi_i}\omega_{\mm}$ when, additionally, $\chi_i\!\to\!0$ ($\omega_3$ and $\omega_2$ replace $\omega_{2b}$ and $\omega_{2f}$, respectively, in the overdamped region). Consequently, $\epsilon_3$ and $\epsilon_2$, purely real and negative, behave in the limit as $\epsilon_3\!\to\!0^{\mm}$ and $\epsilon_2\!\to\!-(\frac{\chi_i}{(\chi_r+1)\chi_r})^2\epsilon_{\mm}$ (see panel (a2)). Finally, we have $x_3\!\to\!0$ and $x_2\!\to\!f_2\!+\!b_2\!=\!\frac{\epsilon_1-\epsilon_{\mm}}{\epsilon_1}$ (panel (b)).

For a given $\omega_0/\omega_{\mm}$ ratio, $\omega_3$ ($\omega_2$) is directly (inversely) proportional to $\chi_i$, which means, in principle, that the oscillation will die out faster (slower) as we increase loss. However, $x_3\!\to\!0$ in the limit $\omega_0/\omega_{\mm}\!\to\!\infty$, so all we care about is $\omega_2$, which is indeed bounded by $i\frac{(\chi_r+1)\chi_r}{\chi_i}\omega_{\mm}$. That is, the larger the loss, the slower the non-oscillatory damping, which is perfectly consistent with intuition: we need loss to be infinitesimally small in order to make non-oscillatory damping instantaneous right after the temporal discontinuity; this is better understood by noting that, in our circuital analogy, $L\!\to\!0$ in the limit $\omega_0/\omega_{\mm}\!\to\!\infty$, so the RC time constant dictates the decay rate (note that this is the opposite of the underdamped regime, where the damped oscillation from the pair $(\omega_{2f},\omega_{2b})$ will die out faster as we increase loss). This is illustrated in panel (c), where the electric fields decay one order of magnitude faster for $\chi_i\!=\!10^{-4}$ (markers) than for $\chi_i\!=\!10^{-3}$ (lines). 

Moreover, note that the normalized frequencies $\omega_{2f}\chi_i/\omega_{\mm}$ and $\omega_{2b}\chi_i/\omega_{\mm}$ do not depend on $\chi_i$ when plotted vs. $\omega_0\chi_i/\omega_{\mm}$---as shown in panel (a1) of Fig.~\ref{fig:C1}, where lines and markers represent different values of $\chi_i$---, very much like the normalized dielectric functions in panel (a2) and the amplitude coefficients in panel (b). Interestingly, we know from Sec.~\ref{sec:lossy} that, at the critical point, both $f_2$ and $b_2$ diverge, though with bounded $f_2\!+\!b_2$: not only do we now observe this behavior, but also $f_2\!+\!b_2$ remains constant (see magenta and orange plots in panel (b)). 

\begin{figure}[h]
\includegraphics[width=3.4in]{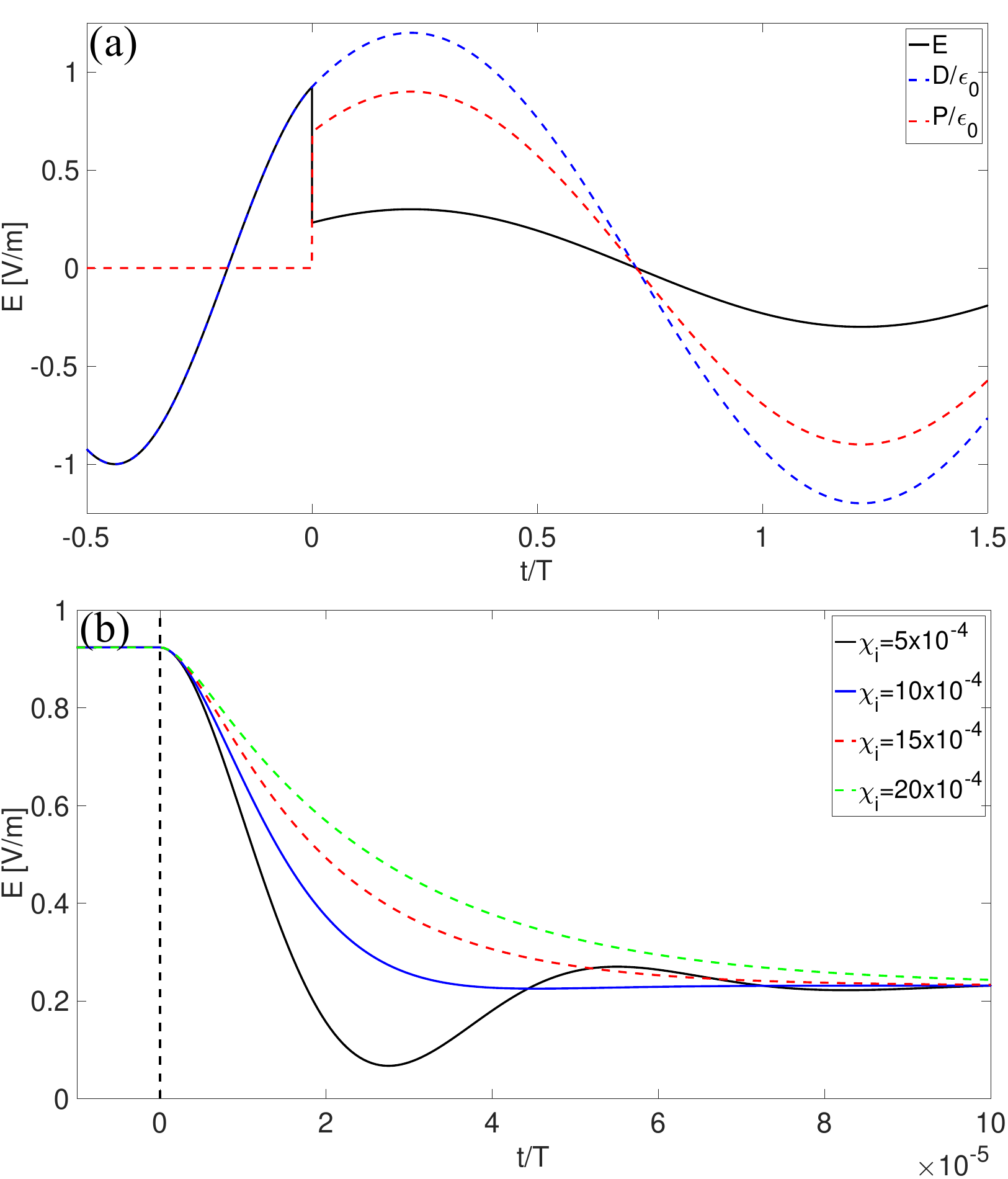}
\caption{(a) Electromagnetic waves vs. time at $z\!=\!\lambda/16$ for the transition from vacuum to $\epsilon_{\protect\pp}(\omega_{\protect\mm})\!=\!4\!-\!i\chi_i$, considering $\omega_0\!=\!10^4\omega_{\protect\mm}$ and several values of $\chi_i$ specified in panel (b); all the curves are virtually the same and overlap with the chosen $t$-axis scale, converging to the solution of the nondispersive lossless case. (b) Zoomed-in view of the transition of panel (a), both for underdamped (solid lines) and overdamped (dashed lines) scenarios.}
\label{fig:C2}
\end{figure}

Finally, in Fig.~\ref{fig:C2} we show how, in the limit of $\chi_i\!\to\!0$ and $\omega_0\!\to\!\infty$, the resulting waves, though continuous, converge to the well-known solution of a nondispersive medium undergoing a step-like change in its dielectric function \cite{1124533,Xiao:14,8858032}, with discontinuous $E$ and $P$ (black and red lines in panel (a), respectively). 

\section{Satisfaction of Parseval's Theorem}
We herein show how one can still find a form of the Parseval–Plancherel theorem \cite{parseval1806memoire,plancherel1910contribution}---also known as Rayleigh's energy theorem \cite{rayleigh1889liii})---that is satisfied by our infinite-energy double-sided signals. Assuming a lossless Lorentzian, $E_{\pp}(t)$ in Eq.~\eqref{eq:2B1a} will have infinite energy, and yet we can consider some positive real number $\sigma$ such that $e^{-\sigma t}E_{\pp}(t)$ is Lebesgue square-integrable \cite{lebesgue1902integrale}: $e^{-\sigma t}E_{\pp}(t)\in L^2(0,\infty)$. Direct application of Parseval's theorem for finite-energy signals results in
\begin{equation}
\begin{split}
\int_{0}^{\infty}\!\!\!\!e^{-2\sigma t}\vert E_{\pp}(t)\vert^2dt&=\frac{1}{2\pi}\!\!\int_{-\infty}^{\infty}\!\!\!\!\vert \mathcal{F}\mathcal{T}\{e^{-\sigma t}E_{\pp}(t)U(t)\}(\omega)\vert^2d\omega \\
&=\frac{1}{2\pi}\!\!\int_{-\infty}^{\infty}\!\!\!\!\vert \mathcal{L}_{r}\{E_{\pp}(t)\}(\sigma+i\omega)\vert^2d\omega,\label{eq:C1}
\end{split}
\end{equation}
with $\mathcal{L}_{r}\{\}$ the unilateral (right-sided) Laplace transform. Similar considerations allow us to write, for the left-sided signal $E_{\mm}$,
\begin{equation}
\begin{split}
\int_{-\infty}^{0}\!\!\!\!e^{2\sigma t}\vert E_{\mm}(t)\vert^2dt&=\frac{1}{2\pi}\!\!\int_{-\infty}^{\infty}\vert \mathcal{F}\mathcal{T}\{e^{\sigma t}E_{\mm}(t)U(-t)\}(\omega)\vert^2d\omega \\
&=\frac{1}{2\pi}\!\!\int_{-\infty}^{\infty}\vert \mathcal{L}_{l}\{E_{\mm}(t)\}(-\sigma+i\omega)\vert^2d\omega.\label{eq:C2}
\end{split}
\end{equation}

Finally, we can write, for our double-sided signal $E(t)\!=\!E_{\mm}U(-t)\!+\!E_{\pp}U(t)$, the energy equality
\begin{equation}
\begin{split}
\int_{-\infty}^{\infty}e^{-2\sigma \vert t\vert}\vert E(t)\vert^2dt=\frac{1}{2\pi}\int_{-\infty}^{\infty}\vert \mathcal{F}\mathcal{T}\{e^{-\vert\sigma\vert t}E(t)\}(\omega)\vert^2d\omega \\
=\frac{1}{2\pi}\int_{-\infty}^{\infty}\vert \mathcal{L}_{l}\{E(t)\}(-\sigma+i\omega)+\mathcal{L}_{r}\{E(t)\}(\sigma+i\omega)\vert^2d\omega,\label{eq:C3}
\end{split}
\end{equation}
where $(\mathcal{L}_{l}\!+\!\mathcal{L}_{r})\{E(t)\}(s)$ represents the bilateral Laplace transform of $E(t)$, whose region of convergence (ROC) is given in this case by $\vert \text{Re}(s)\vert\!<\!\sigma$.

\nocite{*}

\bibliography{apssamp}


\end{document}